\documentstyle[floats,amsfonts,prd,aps,epsf]{revtex}

\def\mathrm{\rm}
\def\mathbf{\bf}

\newcommand{\insertfigure}[3]{%
\begin{figure}[htbp]%
  \Size#1%
  \begin{center}%
  \begin{minipage}[t]{\Size}%
    \idl%
    \epsfxsize=\Size%
    \epsfbox{#2}%
    \caption{#3}%
  \end{minipage}%
  \end{center}%
\end{figure}%
}

\newenvironment{equation*}{\begin{displaymath}}{\end{displaymath}}
\newcommand{\idl}{\hspace*{-0.15\Size}}

\newtheorem{Def}{Definition}
\newtheorem{Satz}{Satz}
\newtheorem{Theorem}[Satz]{Theorem}

\newcommand{\const}{\mbox{const} }
\newcommand{\Scri}{\mbox{$\cal J$}}

\newlength{\Size}

\newcommand{\NENNER}{{\left(1-\kappa\,
                        \left(\Omega\,\phi/2\right)^2\right)}}

\newcommand{\ei}{ {e_{\f{0}}{}^0} }

\newcommand{\gamP}{\gamma1}
\newcommand{\gamii}{\gamma2}
\newcommand{\gamM}{\gamma3}
\newcommand{\hRi}{\widehat{R1}}
\newcommand{\hRii}{\widehat{R2}}
\newcommand{\hRiii}{\widehat{R3}}
\newcommand{\RICCI}{R}
\newcommand{\DoR}{\left(\partial_{\f{0}}R\right)}
\newcommand{\DiR}{\left(\partial_{\f{1}}R\right)}
\newcommand{\dioio}{d_{\f{1}\f{0}\f{1}}{}^{\f{0}}}

\newcommand{\Omo}{{\Omega_{\f{0}}}}
\newcommand{\Omi}{{\Omega_{\f{1}}}}

\newcommand{\pho}{{\phi_{\f{0}}}}
\newcommand{\phI}{{\phi_{\f{1}}}}
\newcommand{\hphi}{{\widehat{\phi1}}}
\newcommand{\hphiii}{{\widehat{\phi3}}}

\newcommand{\B}[1]{\bar{#1}}
\newcommand{\I}[1]{{}_{\mathrm\mathbf #1 \,}{}}
\newcommand{\f}[1]{\underline{#1}}

\newcommand{\hR}{\mbox{$\hat{R}$}}
\newcommand{\hp}{\mbox{$\hat{\phi}$}}
\newcommand{\hT}{\mbox{$\hat{T}$}}

\newcommand{\tG}{\mbox{$\tilde{G}$}}
\newcommand{\tT}{\mbox{$\tilde{T}$}}
\newcommand{\tg}{\mbox{$\tilde{g}$}}
\newcommand{\tn}{\mbox{$\tilde{\nabla}$}}
\newcommand{\tp}{\mbox{$\tilde{\phi}$}}
\newcommand{\tR}{\mbox{$\tilde{R}$}}

\newcommand{\N}[1]{ {\cal N}\I{#1} }

\newcommand{\RII}{\mbox{$R2$}}

\begin{document}
 
\draft

\title{A Method for Calculating the Structure of (Singular) Spacetimes
 in the Large}

\maketitle

\author{Peter H\"ubner\footnotemark[1]}
\newcommand{\oldthefootnote}{\thefootnote}
\renewcommand{\thefootnote}{\fnsymbol{footnote}}
\footnotetext[1]{This work is to a large extent part of my Ph.~D.\
      thesis which has been done at the Max-Planck-Institut f\"ur
      Astrophysik, Postfach 1523, D-85740 Garching}
\renewcommand{\thefootnote}{\arabic{footnote}}

\address{Max-Planck-Gesellschaft Arbeitsgruppe Gravitationstheorie, an
  der Universit\"at Jena, Max-Wien-Platz 1, D-07743 Jena, 
  (pth@gravi.physik.uni-jena.de)}

\begin{abstract}
A formalism and its numerical implementation is presented which
allows to calculate quantities determining the spacetime structure in
the large directly. This is achieved by conformal techniques by which
future null infinity ($\Scri{}^+$) and future timelike 
infinity ($i^+$) are mapped to grid points on the
numerical grid. The determination of the causal structure of
singularities, the localization of event horizons, the extraction of
radiation, and the avoidance of unphysical reflections at the outer
boundary of the grid, are demonstrated with calculations of
spherically symmetric models with a scalar field as matter and
radiation model. 
\end{abstract}

\pacs{04.20.Dw 04.20.Ha 04.25.Dm}
\section{Introduction}
\subsection{General framework}
From the singularity theorems by S.~Hawking and R.~Penrose it is
known that the appearance of singularities in general relativity is an
unavoidable feature for ``strong'' initial data \cite{HaE73TL}.
Conversely, it has been proven recently that for  
small initial data the future of the initial value hypersurface looks
in the large in a certain sense like the future of flat space data
\cite{ChK93TG,Fr91ot,HuXXgr}.  
Unfortunately, it has turned out to be extremely difficult to obtain
reasonable conjectures --- not to mention the proofs --- about the
properties of a spacetime given by an initial value problem, as illustrated,
for example, by the history of the cosmic censorship hypothesis. 
\\
Getting an overview for the phenomena appearing and 
making reasonable conjectures is an area numerical relativity can
contribute a lot to 
and actually already has; most impressive illustrated by the
discovery of the echoing property by M. Choptuik \cite{Ch93ua}. 
But
most of the codes used so far are designed to analyze local behaviour.
Statements about the behaviour in the large, global issues, are
obtained by assuming that the extension of the grid used approximates
an infinite grid well enough.
This way it is very difficult to get a reliable error estimate and to
decide, what is sufficiently far,  if not impossible at all.
\\
But what global issues are of  interest and why are they interesting?
\\
Firstly, there are questions related to cosmic censorship,
especially the location and penetration of an event horizon, which is
the boundary of the region of spacetime from where no
null geodesic reaches null infinity. In a recent review article
\cite{Cl94ar} about cosmic censorship C.~Clarke writes about the
location of event horizons: ``In terms of
numerical simulations, this means that it is essential to perform the
simulations in the compactified picture in which $\Scri^+$ is
represented explicitly''.
\\
Secondly, there is the whole topic of gravitational radiation on
asymptotically flat spacetimes. Due to the gauge freedom and the
non-linearity of the theory the classification of radiation as ingoing
and outgoing is a difficult issue. In- and outgoing gravitational waves
are defined with respect to null infinity only. The related
difficulties in extracting  
gravitational waves from the grid at finite distances and the problem
with the avoidance 
of an unphysical reflection on the outer boundary of the grid are well

known and
have been a topic for research for a long time. In most
methods developed so far the error made consists of the error from
reading off at finite distances and the discretization error.
\\
In this paper I will present the numerical implementation of a
formalism allowing to calculate a ``compactified'' spacetime
including $\Scri^+$ and $i^+$. The solution of the problems
concerning radiation is trivial by construction as will be seen. The
only errors which 
may appear are caused by the discretization error of the numerical
partial differential equation solver. These errors can be estimated
and controlled by grid extrapolation techniques (e.~g.\ Richardson
extrapolation). Although the calculations have been done
under the assumption of spherical symmetry with a scalar field as model for
radiation the simplicity and exactness of radiation extraction hold
for arbitrary symmetry assumptions. 
\\
Furthermore, due to the spherical
symmetry a special coordinate gauge could be found allowing to
cover the complete domain of dependence of the initial hypersurface
and to calculate the causal structure of a singularity. The location
of the event horizon is straightforward.
\\
The formalism is based on conformal techniques developed by R. Penrose
to describe asymptotically flat spacetimes. By a rescaling
$g_{ab}=\Omega^2\,\tg_{ab}$ the physical spacetime $(\tilde
M,\tg_{ab})$ is mapped to an unphysical spacetime 
$(M,g_{ab},\Omega{})$ with boundary \Scri{}. The boundary represents
null infinity. $M$ is a ``compactified''\footnote{$M$ is not really
  compact --- in Minkowski spacetime there are three points, future
  and past timelike and spacelike infinity, missing. In general $M$
  cannot be smoothly extended to contain those points.}  version of
$\tilde M$. Gravitational 
radiation is the value of certain components of the Weyl spinor on
that boundary. In the originally suggested form the formalism is not 
suited to describe initial value problems. H. Friedrich has modified
it to describe initial value problems --- one has to solve a set of
evolution equations for the unphysical spacetime
$(M,g_{ab},\Omega{})$. The rescaling factor $\Omega{}$ acts like an
artificial matter field for the Einstein equations. Subsection
\ref{MathForm} reviews the equations for the unphysical spacetime.
\\
J. Winicour and R. Gomez have
developed an approach which uses Bondi's ideas for describing
gravitational radiation: Outgoing null cones, with the area distance
and the direction angle as coordinates on it, are compactified to
represent future null infinity by grid points. Their method gives
simpler equations but has certain disadvantages: The existence of smooth 
outgoing null cones is essential. Null caustics, which are caused
e.g.~by gravitational lensing effects,  must not appear. Furthermore the
evolution scheme cannot penetrate event horizons. 
The extraction of radiation is complicated slightly as the Weyl spinor is
not a variable of the system --- the determination of
gravitational radiation requires to calculate numerically derivatives
on \Scri{}. Their formalism solves a characteristic initial value, the
formalism presented here solves a ``normal'', spacelike initial value
problems in unphysical spacetime.
\subsection{Description of asymptotically flat spacetimes with
  conformal techniques}
Shortly after H.~Bondi et.~al. \cite{BoBA62gw} proved that
gravitational radiation is not a gauge effect R.~Penrose suggested a
coordinate independent way to characterize asymptotically flat spaces
and gravitational radiation. A thorough discussion
of the ideas and the interpretation can be found at various places in
the literature, e.g.\ \cite{Ge76as,Pe64ct}. The
definitions of asymptotical flatness given in the literature differ
slightly. The following will be used here:
\begin{Def}
\label{asymFlat}
  A spacetime $(\tilde M, \tg_{ab})$ is called {\bf asymptotically
    flat} if there is another ``unphysical'' spacetime $(M,g_{ab})$
  with boundary \Scri{} and a smooth embedding by which $\tilde M$
  can be identified with $M-\Scri{}$ such that:
  \begin{enumerate}
  \item There is a smooth function $\Omega$ on $M$ with
    \begin{equation*}
      \Omega \mid_{\tilde M} > 0 \qquad \mbox{and} \qquad 
      g_{ab} \mid_{\tilde M} = \Omega^2 \tg_{ab}.
    \end{equation*}
  \item On \Scri{} 
    \begin{equation*}
      \Omega = 0 \qquad \mbox{and} \qquad \nabla_a \Omega \ne 0.
    \end{equation*}
  \item \label{nullCompleteness} Each null geodesic in $(\tilde
    M,\tilde g_{ab})$ acquires a past and a future endpoint on \Scri{}.
  \end{enumerate}
\end{Def}
Because of item~\ref{nullCompleteness} null geodesically incomplete
spacetimes like Schwarzschild are not asymptotically flat. The next
definition includes those spacetimes which have only an
asymptotically flat part:
\begin{Def}
\label{weakasymFlat}
  A spacetime is called {\bf weakly asymptotically flat} if
  definition~\ref{asymFlat} with the exception
  of item~\ref{nullCompleteness} is fulfilled. 
\end{Def}
Definition \ref{asymFlat} and \ref{weakasymFlat} classify spacetimes,
they do not require that the Einstein equation is fulfilled. For a
physical problem one would like to give ``asymptotically flat data''
and evolve them according to the Einstein equation. 
\\
Nevertheless the geometrically description was extremely helpful in
analyzing asymptotically flat spacetimes and it has been successfully
used as guideline to construct a formalism for
analyzing initial value problems. This method has been developed and
applied to various matter sources by H.~Friedrich~
\cite{Fr91ot,Fr81on,Fr83cp,Fr85ot,Fr86op,Fr86ot,Fr88os} and myself
\cite{HuXXgr}.
\\
The idea is to choose a spacelike initial value surface in the
unphysical spacetime $(M,g_{ab})$ and to evolve it.
\\
For Minkowski space the unphysical spacetime $(M,g_{ab})$ can be
smoothly extended with three points, future $(i^+)$ and past  $(i^-)$
timelike infinity, the end and the starting point of all
timelike geodesics of $(\tilde M,\tg_{ab})$ respectively, and
spacelike infinity $(i^0)$, the 
end point of all spacelike geodesics of $(\tilde M,\tg_{ab})$. The 
point $i^0$
divides \Scri{} into two disjunct parts, future $(\Scri^+)$ and past
$(\Scri^-)$ null infinity. It is well known and has been discussed
elsewhere that there are unsolved problems in smoothly
extending a ``normal'' Cauchy hypersurface of $\tilde M$ to $i^0$ if the
spacetime has non-vanishing ADM mass. Certain curvature quantities
blow up at $i^0$, reflecting the non-invariance of the mass under
rescalings. 
\\
By choosing a spacelike (with respect to $g_{ab}$) hypersurface $S$
not intersecting $i^0$ but $\Scri^+$ ($\Scri^-$) we avoid the problems
with $i^0$. $S$ is called a hyperboloidal hypersurface --- the
corresponding initial value problem is called a hyperboloidal initial
value problem. 
\\
The well-posedness of the hyperboloidal initial value problem for
general relativistic scalar fields has been discussed in
\cite{HuXXgr}. There, a precise definition of a hyperboloidal initial
value problem can be found, too.
\\
Figure \ref{RegUnRZ} shows a diagram of the unphysical spacetime with
an example of a hyperboloidal surface in it. In figure \ref{RegRZ} the
corresponding physical spacetime is shown. In both figures only one
space coordinate is drawn. All points, except those on the axis and
$i^0$, represent spheres. 
\insertfigure{7cm}{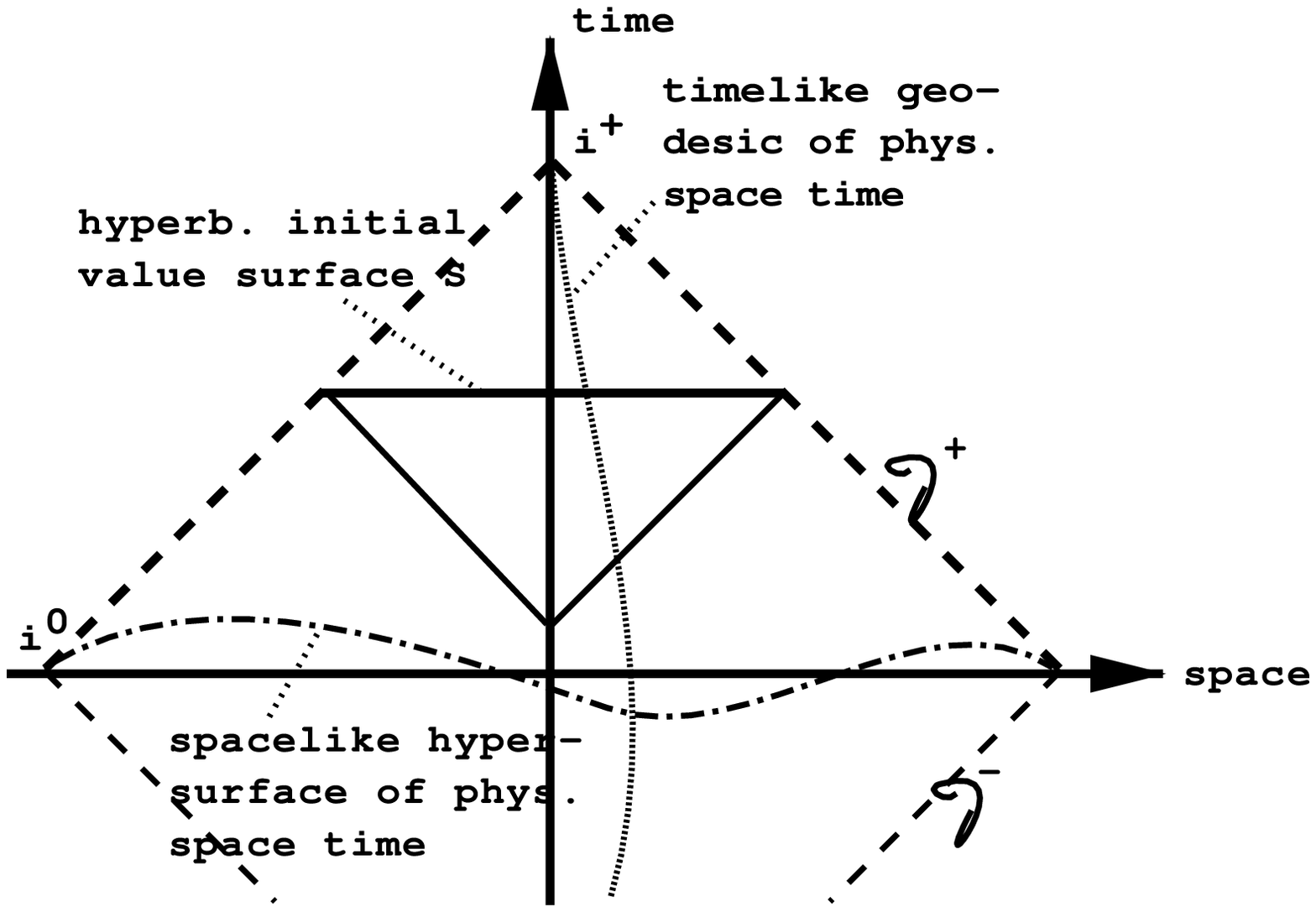}{
    \label{RegUnRZ}Unphysical Minkowski spacetime}  
\insertfigure{7cm}{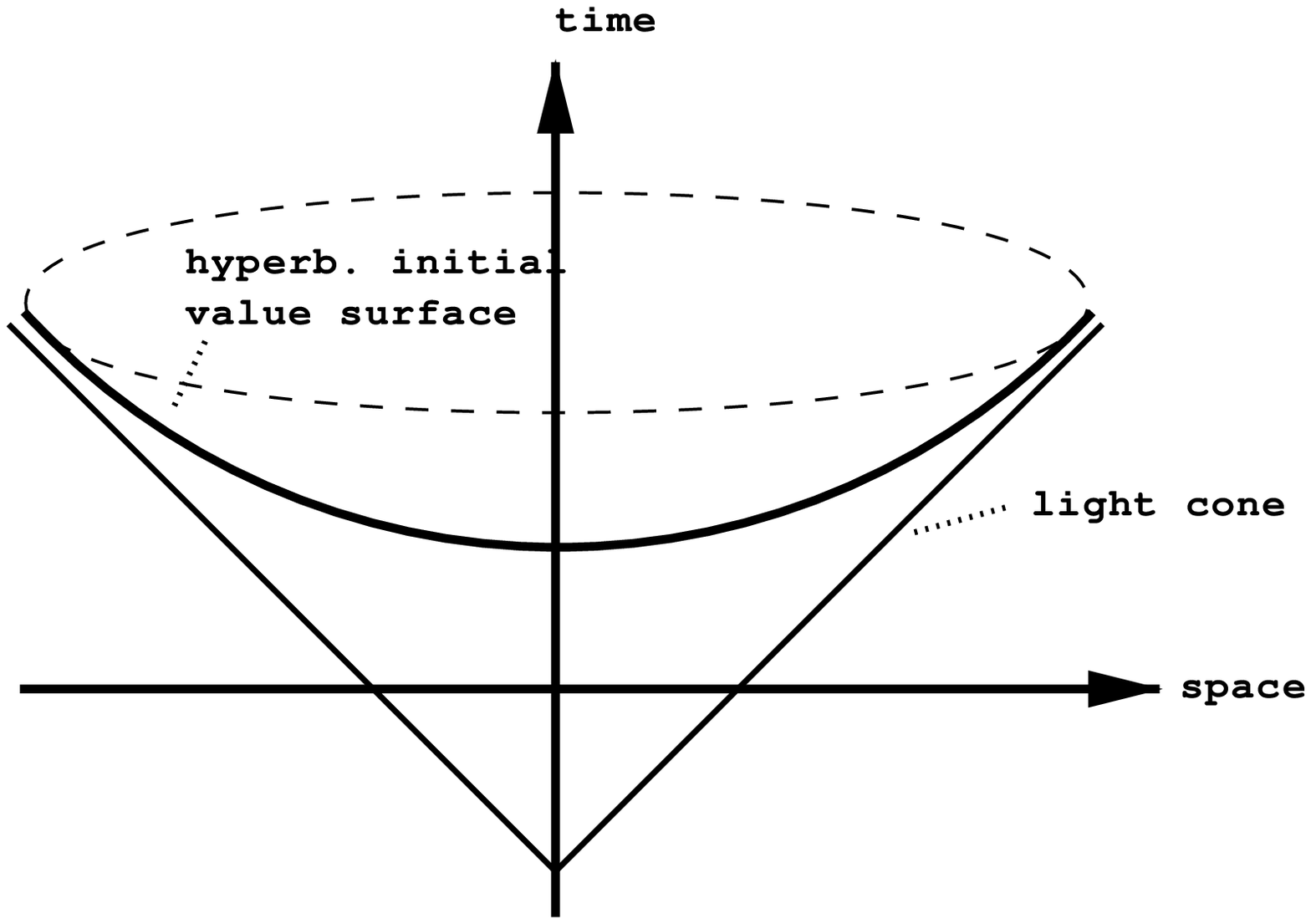}{
    \label{RegRZ}Physical Minkowski spacetime} 
The domain of dependence $D(S)$ of 
$S$ will not contain the complete spacetime. The interior of $S$
corresponds to an everywhere spacelike hypersurface in the physical
spacetime which approaches a null hypersurface $N$ asymptotically. If $N$
is a light cone $L$ then the domain of dependence of $S$ is $L$. Therefore
the hyperboloidal initial value problem is well suited to describe the
future (past) of data on the spacelike hypersurface $S$, e.~g.~a
stellar object and the gravitational radiation caused by its future
time evolution. It is not well suited to investigate the structure
near $i^0$.
\\
Why is it advantageous to solve the initial value problem in the
unphysical spacetime? The future of
the data in physical spacetime $(\tilde M,\tg_{ab})$ is completely
determined by the data on the interior of $S$. Since the rescaling
factor $\Omega{}$ is known, the calculation of quantities of physical
spacetime from the unphysical quantities is merely algebra. In
unphysical spacetime $\Scri^+$ is on the grid. $\Scri{}^+$ is an
ingoing null cone 
starting at the 2-surface in $S$ where $\Omega{}$ vanishes. By
extending $S$ a little bit beyond the intersection with \Scri{}
in unphysical spacetime we do not change the physical spacetime
$(\tilde M,\tg_{ab})$. But with a finite number of gridpoints and thus
a finite computation time the whole future of the initial hypersurface
is covered.  And since $\Scri{}^+$ is an ingoing null line the values
at $\tilde M$ and 
$\Scri{}^+$  --- describing the physics --- do not depend on the
values at the outer boundary of the extended $S$. The numerical
dependence on the outer boundary must converge to $0$.  
\\
Future null infinity, $\Scri{}^+$, which can be found by searching for
the $\Omega=0$ contour, 
is on the grid. By an appropriate choice of gauge it can even 
be arranged that the position of \Scri{} on the grid is known by
analytical considerations. In the worst case, one has to interpolate
between neighbouring grid points to find the radiation emitted. In
the case of gravitational radiation, it is given by certain
components of the Weyl tensor, which are variables in the system of
equations for the unphysical spacetime.
\section{The model and the system of equations in unphysical spacetime}
\subsection{The model}
The choice of the model is always a compromise between generality and
manageability. For this work it was necessary to have some model for
radiation to demonstrate the simplicity of the radiation extraction in
the formalism. On the other side the spacetime symmetry should be as
high as possible to reduce the demand for computational resources and
to simplify coordinate choice related questions. 
\\
The spacetime is assumed to be spherically
symmetric, the radiational degree of freedom is modeled by a
conformally invariant scalar field. The equations are 
\begin{mathletters}
\label{model}
\begin{eqnarray}
\label{Wllngl}
  \tilde{\vphantom{\phi}\Box} \tp - \frac{\tR}{6} \, \tp & = & 0
\\
\label{EinstPhys}
  ( 1 - \frac{1}{4} \kappa \,\tp^2 ) \, \tR_{ab} & = &
       \left(
          \kappa \, (\tn_a \tp) (\tn_b \tp) 
          - \frac{1}{2} \, \kappa \, \tp \, \tn_a \tn_b \tp
          - \frac{1}{4} \, \kappa \, \tg_{ab} (\tn^c \tp) (\tn_c \tp) 
       \right),
\end{eqnarray}
where the second is equivalent to $\tG_{ab} = \kappa \, \tT_{ab}$ with  
\begin{equation}
\label{TkonfS}
  \tT_{ab} = (\tn_a \tp) (\tn_b \tp) - \frac{1}{2} \, \tp \, 
\tn_a \tn_b \tp
        + \frac{1}{4} \, \tp^2 \tR_{ab} -
        \frac{1}{4} \, \tg_{ab} 
          \left( (\tn^c \tp) (\tn_c \tp) + \frac{1}{6} \, \tp^2 \tR
          \right).
\end{equation}
\end{mathletters}
The notation used is explained in appendix \ref{Konventionen}.
\\
The form invariance of (\ref{Wllngl}) under the rescalings $g_{ab} =
\Omega^2 \, \tg_{ab}$ and $\phi = \Omega^{-1} \, \tp{}$ was the reason
for not choosing the massless Klein-Gordon scalar field with its simpler
equations in physical spacetime. In \cite{HuXXgr} it has been
shown that the initial value problems for these matter models are
equivalent for practical purposes. 
\subsection{The equations for unphysical spacetime}
\label{MathForm}
In \cite{HuXXgr} it has been argued that the Einstein equation, if
simply translated to the unphysical spacetime, becomes
``singular'' on \Scri{} and thus is not suitable to calculate the
unphysical spacetime $(M,g_{ab})$. For a derivation of regular
equations for the unphysical spacetime the reader is referred to
\cite{HuXXgr}, here only the final first order set of equations is
given. 
\\
In this first order system, the variables are the components of the
frame $e_{\f{k}}{}^a$ with respect to the coordinates $x^\mu{}$, 
$e_{\f{k}}{}^\mu{}$, 
the frame components $\gamma^{\f{m}}{}_{\f{j}\f{k}}$ of the Ricci
rotation coefficients $\gamma^a{}_{\f{j}\f{k}}$, certain combinations of the
components of the tracefree part $\hR_{ab}$ of the Ricci tensor and
the regularized Weyl tensor $d_{abc}{}^d := \Omega^{-1} \, C_{abc}{}^d$,   
the conformal factor $\Omega{}$, the frame components $\Omega_{\f{i}}$
of its gradient $\Omega_a$, the trace of its 
second derivative $\omega{}$, the conformal scalar field $\phi{}$, the
frame components $\phi_{\f{i}}$ of its gradient $\phi_a$ and
combinations of its second derivatives $\phi_{ab} =: \hat\phi_{ab} +
\frac{1}{4} \, g_{ab} \, \phi_c{}^c$. The first order system, written
in abstract index notation is:
\begin{mathletters}
\label{quaSys}
\begin{eqnarray}
\label{Ne}
  \N{e}^a{}_{bc} & := & T^a{}_{bc}   = 0 \\
\label{Ng}
  \N{\gamma}_{abc}{}^d & := &
    R\I{diff}_{abc}{}^d - R\I{alg}_{abc}{}^d  = 0 \\  
\label{NR}
  \N{R}_{abc} & := &
    \nabla_{[a} \hR_{b]c} 
    + \frac{1}{12} (\nabla_{[a} R) \, g_{b]c} - \Omega_d \, d_{abc}{}^d
    + \Omega \, m_{abc} 
    - \frac{2}{3} \, \Omega \, m_{[a|d|}{}^d g_{b]c} \nonumber \\
    & = & 0 \\
\label{Nd}
  \N{d}_{abc} & := &
  \nabla_d d_{abc}{}^d  - m_{abc} + \frac{2}{3} \, m_{[a|d|}{}^d \, g_{b]c}
  = 0 \\
\label{NO}
  \N{\Omega}_a & := & \nabla_a \Omega - \Omega_a = 0              \\
\label{NDO}
  \N{D\Omega}_{ab} & := &
    \nabla_a \Omega_b + \frac{1}{2} \, \Omega \, \hR_{ab} 
    - \omega \, g_{ab} - \frac{1}{2} \, \Omega^3 \, T_{ab} = 0           \\
\label{No}
  \N{\omega}_{ab} & := &
    \nabla_a \omega + \frac{1}{2} \, \hR_{ab} \, \Omega^b
    + \frac{1}{12} \, R \, \Omega_a + \frac{1}{24} \, \Omega \, \nabla_a R
    - \frac{1}{2} \, \Omega^2 \, T_{ab} \, \Omega^b   \nonumber  \\
    & = & 0 \\
\label{Np}
  \N{\phi}_a & := & \nabla_a \phi - \phi_a = 0
\\
\label{NDp}
  \N{D\phi}_{ab} & := & 
  \nabla_a \phi_b - \hat{\phi}_{ab} - \frac{1}{4} \phi_c{}^c \, g_{ab} = 0
\\
\label{NBp}
  \N{\Box\phi} & := & \phi_a{}^a - \frac{R}{6} \, \phi = 0
\\
\label{NDDp}
  \N{DD\phi}_{abc} & := & 
  \nabla _{[a} \hat{\phi}_{b]c} 
  + \frac{1}{6} \, ( \phi \, \nabla_{[a} R + R \, \phi_{[a} ) \, g_{b]c}
  - \frac{1}{2} \, R\I{alg}_{abc}{}^d \, \phi_d = 0
\\
\label{NDBp}
  \N{D\Box\phi}_a & := &
  \nabla _{a} \phi_b{}^b 
  - \frac{1}{6} \, ( \phi\, \nabla_a R + R \, \phi_{a} ) = 0,
\end{eqnarray}
\end{mathletters}
where $T^a{}_{bc}$ is the torsion whose components are given by 
\begin{equation*}
    T^{\f{i}}{}_{\f{j}\f{k}} = 
    \left( e_{\f{j}}(e_{\f{k}}{}^\mu) -
           e_{\f{k}}(e_{\f{j}}{}^\mu) \right) 
      e^{\f{i}}{}_\mu 
  + \gamma^{\f{i}}{}_{\f{j}\f{k}} - \gamma^{\f{i}}{}_{\f{k}\f{j}}
\end{equation*}
the tensor $R\I{alg}_{abc}{}^d$ is an abbreviation for
\begin{equation}
\label{Ralg}
  R\I{alg}_{abcd} = 
  \Omega \, d_{abcd} 
  + g_{c[a} \hR_{b]d} - g_{d[a} \hR_{b]c}
  + \frac{1}{6} \, g_{c[a} \, g_{b]d} \, R,
\end{equation}
$R\I{diff}_{abc}{}^d$ has the components 
\begin{eqnarray*}
    R\I{diff}_{\f{i}\f{j}\f{k}}{}^{\f{l}} & = &
  e_{\f{j}}(\gamma^{\f{l}}{}_{\f{i}\f{k}}) 
    - e_{\f{i}}(\gamma^{\f{l}}{}_{\f{j}\f{k}}) 
  - \gamma^{\f{l}}{}_{\f{i}\f{m}} \, \gamma^{\f{m}}{}_{\f{j}\f{k}} 
  + \gamma^{\f{l}}{}_{\f{j}\f{m}} \, \gamma^{\f{m}}{}_{\f{i}\f{k}}
  \nonumber  \\
  && \qquad
  + \gamma^{\f{m}}{}_{\f{i}\f{j}} \, \gamma^{\f{l}}{}_{\f{m}\f{k}} 
  + \gamma^{\f{m}}{}_{\f{j}\f{i}} \, \gamma^{\f{l}}{}_{\f{m}\f{k}}
  - \gamma^{\f{l}}{}_{\f{m}\f{k}} \, T^{\f{m}}{}_{\f{j}\f{i}},
\end{eqnarray*}
the ``energy momentum tensor'' in unphysical spacetime is
\begin{equation*}
  T_{ab} = 
                 \phi_a \, \phi_b
                  - \frac{1}{2} \, \phi \, \phi_{ab}
                  + \frac{1}{4} \, \phi^2 \, R_{ab} 
                  - \frac{1}{4} \, g_{ab} 
                  \left( \phi_c\, \phi^c
                         + \frac{1}{6} \, \phi^2 \, R \right)
\end{equation*}
and 
\begin{eqnarray*}
  \lefteqn{ m_{abc} = \frac{1}{1-\frac{1}{4} \, 
\Omega^2\phi^2} \qquad * }   \\ 
  \lefteqn{ \Big( \, \Omega } && \qquad
  \big[ \frac{3}{2}  \, \phi_{[a} \phi_{b]c}
    - \frac{1}{2}  \, g_{c[a} \phi_{b]d} \phi^d 
    + \frac{1}{4}  \, \phi  \, \Omega d_{abc}{}^d \phi_d
    + \frac{1}{4}  \, \phi  \, g_{c[a} \hR_{b]}{}^d \phi_d
    - \frac{3}{4}  \, \phi  \, \phi_{[a} \hR_{b]c}              
    \\ && \qquad \quad
    - \frac{1}{12}  \, \phi  \, \phi_{[a} g_{b]c} R
    + \frac{1}{4}  \, \Omega  \, \phi^2 d_{abc}{}^d \Omega_d  \big] 
    \\ && \quad
  - 3  \, \Omega_{[a} \big[ \phi_{b]} \phi_c
    - \frac{1}{2}  \, \phi  \, \phi_{b]c}
    + \frac{1}{4}  \, \phi^2 \hR_{b]c}
    + \frac{1}{36}  \, \phi^2 g_{b]c} R
    - \frac{1}{3}  \, g_{b]c}  \, \phi^d \phi_d \big]             
      \\ && \quad
  + \Omega^d g_{c[a} \big[ \phi_{b]} \, \phi_d
    - \frac{1}{2}  \, \phi  \, \phi_{bd}
    + \frac{1}{4}  \, \phi^2 \hR_{b]d}    \big]     \quad     \Big).
\end{eqnarray*} 
Unfortunately the terms introduced into the system by the scalar
field, $T_{ab}$ and $m_{abc}$, 
complicate the equations a lot due to the form of the energy
momentum tensor for the conformally invariant scalar field. 
\\
Furthermore there is an additional equation,
\begin{equation}
\label{SpGl}
  \Omega^2 \, R + 6 \, \Omega \, \nabla^a \nabla_a \Omega 
    - 12 \, (\nabla^a \Omega) \, (\nabla_a \Omega)  
  = 0,
\end{equation}
which must be satisfied at one point to be automatically fulfilled
everywhere. 
\\
If we fulfill equation (\ref{SpGl}) at one point the function $R(t,r)$ 
can be given freely. It determines the gauge freedom in the conformal
factor $\Omega{}$ to a certain extent. If $(M,g_{ab},\Omega)$ is a
solution of the unphysical equations so is
$(M,\theta^2\,g_{ab},\theta\,\Omega)$ for $\theta>0$. In all the
calculations, $R$ has been set to $6$, the value obtained if the
compactification for Minkowski space given in \cite{Wa84GR} is used.
The program code does allow to specify $R(t,r)$. The conformal gauge
freedom has been discussed in more detail in \cite{HuXXgr}.  
\subsection{Simplifications by the spherical symmetry and the
  remaining gauge freedom}
The first order system (\ref{quaSys}) is an underdetermined system. To
make it complete the coordinate and frame gauge
freedom must be fix. This will be done in this section. 
\\
Furthermore the number of variables will be reduced 
significantly by making use of the spherical symmetry. 
\\
A spacetime is said to be spherically symmetric if it possesses a
isometry group $I$ which contains a subgroup $R$ which is isomorphic to the
3-dimensional group ${\rm SO}(3)$ and whose orbits are orientable
submanifolds of dimension $< 3$. The orbits of $R$ are either (fix) points
or spheres $\Bbb{S}^2$. The fix points form at most two timelike
geodesics as can be seen by symmetry arguments. It is assumed that the
initial hypersurface intersects with at least one line of fix points,
i.e.~there is a (regular) center on the initial hypersurface. 
\\
But no attempt was made to use every simplification spherical symmetry
provides for the following reason: It is well known that Einstein's
equations can be reduced to two ordinary differential equations
(constraints) \cite{Ch92CB} plus matter equations in spherical
symmetry. The circumstance that no equation with a time derivative for
the geometry variables must be solved 
can be viewed as expressing the lack of gravitational radiation in
spherical symmetry. I have not made any attempt to incorporate this
kind of simplification into the system for the unphysical spacetime,
since the calculations are supposed to be a playground for testing and
learning about the advantages and disadvantages of the formalism for a
later application to models with less symmetry.
\\
In the gauge used the system is symmetric hyperbolic and
the coordinates cover the whole domain of dependence of the initial
hypersurface. It is not known to the author how to construct such a
coordinate system in less symmetric spacetimes. The very nice feature
of semilinearity is certainly an artifact of spherical symmetry. 
\subsubsection{Gaussian gauge}
In the analytical analysis of the initial value problem with
scalar fields as source terms, a Gaussian gauge was used \cite{HuXXgr}.
This gauge gives a symmetric hyperbolic subsidiary system of
equations. If the energy momentum tensor fulfills certain conditions,
the coordinate system will break down because of the formation of
caustics \cite{Wa84GR}. Whether the region of the unphysical spacetime
representing the physical spacetime can be covered by the use of a
Gaussian gauge in unphysical spacetime is not obvious since the
conformal rescaling factor 
$\Omega{}$ acts as a kind of artificial energy-momentum source in
unphysical spacetime. By numerical calculations it was straightforward
to show that even in unphysical spacetime a Gaussian gauge will lead
to caustics \cite{Hu93nu}, and is thus inappropriate to analyze the
strong field regime.
\\
The system of equations is very similar to the one obtained in
``double null'' coordinates except that it is not semilinear. It will
not be discussed any further. 
\subsubsection{``Double null'' gauge}
The construction of the coordinate system and the frame are described
starting from an initial hypersurface. The remarks about the
differentiability class of the objects assume that the construction is
done with respect to a given $C^\infty{}$ manifold (for a definition
see \cite[section 1.1.1]{Wa84GR}). The investigation of the
differentiability is necessary to ensure that a discontinuity of a
value or even a singular value of a variable at the center is not
caused by lacking smoothness of the coordinates. 
\\
The gauge realized in this way turns out to be an obvious adherent of
double null coordinates.  
\\
Assume we are given a spacelike $C^\infty{}$ hypersurface $S$ with
$t=t_0$. This
hypersurface is factored by the orbits of the group. By the geodesics
running from the center in the different directions the angle
coordinates $(\vartheta,\varphi)$ with line element
$d\vartheta^2+\sin^2\vartheta\,d\varphi{}$ on the unit sphere are defined on
all orbits. By isotropy 
the geodesics are perpendicular to the orbits. The orbits
are labeled by a monotonically increasing coordinate $r$, defined to
be $0$ at the center. Auxiliary coordinates $(u,v)$ are defined by
$u:=(t_0-r)/2$ and $v:=(t_0+r)/2$. Set $(u,\vartheta,\varphi)=\const{}$
along the future directed, outgoing null lines,
$(v,\vartheta,\varphi)=\const{}$ along the future directed, ingoing null
line. On passing through the center set $u=v$. Due to the spherical
symmetry there cannot be any null caustic, a break down of the coordinates
is aligned with a spacetime singularity. By $(u,v)$ a timelike coordinate
\begin{equation}
   t=u+v
\end{equation}
and a spacelike coordinate 
\begin{equation}
   r=v-u
\end{equation}
are defined everywhere in the domain of dependence of $S$. An
orthonormal frame field
$(e_{\f{0}}{}^a,e_{\f{1}}{}^a,e_{\f{2}}{}^a,e_{\f{3}}{}^a)$ is defined
except in the center (polar coordinate singularity) 
by normalizing the orthogonal vector fields
$(\partial_t{}^a,\partial_r{}^a,\partial_\vartheta{}^a,
\partial_\varphi{}^a)$. 
\\
As  shown in \cite{Hu93nu} the $t=\const{}$ hypersurfaces are at least
$C^3$ and hence sufficiently smooth. The proof proceeds in two steps,
firstly it is shown that there is a $C^\infty{}$ transformation to
radar coordinates, secondly a theorem by U. Proff 
\cite[theorem 1.2.4]{Pr85rb}
about radar coordinates completes the proof.   
\\
In this coordinate system the following relations hold:
The frame--coordinate matrix is diagonal,
\begin{equation}
  \left( e_{\f{0}}{}^\mu, e_{\f{1}}{}^\mu, 
         e_{\f{2}}{}^\mu, e_{\f{3}}{}^\mu 
  \right) =
  \left(
  \begin{array}{cccc}
    e_{\f{0}}{}^0(t,r) & 0                  & 0                  & 0 \\
    0                  & e_{\f{1}}{}^1(t,r) & 0                  & 0 \\
    0                  & 0                  & e_{\f{2}}{}^2(t,r) & 0 \\
    0                  & 0                  & 0                  & 
      e_{\f{2}}{}^2(t,r)/\sin\vartheta  
  \end{array} 
  \right),   
\end{equation}
and 
\begin{equation}
  e_{\f{0}}{}^0(t,r) = e_{\f{1}}{}^1(t,r).
\end{equation}
All Ricci rotation coefficients except  
\begin{eqnarray}
  && \gamma^{\f{0}}{}_{\f{0}\f{1}}(t,r), \qquad
     \gamma^{\f{0}}{}_{\f{1}\f{1}}(t,r), 
  \nonumber \\
  && \gamma^{\f{0}}{}_{\f{2}\f{2}}(t,r), \qquad
     \gamma^{\f{0}}{}_{\f{3}\f{3}}(t,r) = 
\gamma^{\f{0}}{}_{\f{2}\f{2}}(t,r), 
  \nonumber \\
  && \gamma^{\f{1}}{}_{\f{2}\f{2}}(t,r), \qquad
     \gamma^{\f{1}}{}_{\f{3}\f{3}}(t,r) =
       \gamma^{\f{1}}{}_{\f{2}\f{2}}(t,r),
       \qquad
     \gamma^{\f{2}}{}_{\f{3}\f{3}} = -
        \frac{\cos\vartheta}{\sin\vartheta} \, e_{\f{2}}{}^2
\end{eqnarray}
vanish.
\\
Scalars invariant under rotations are functions of $(t,r)$ only,
rotationally invariant vectors $V^a$ are of the form
\begin{equation}
  V^{\f{i}} = 
    \left( 
    \begin{array}{c}
    V^{\f{0}}(t,r) \\  V^{\f{1}}(t,r) \\ 0 \\ 0
    \end{array}
    \right),
\end{equation}
and symmetric covariant 2-tensors $S_{ab}$, e.g.~the Ricci tensor
$R_{ab}$, invariant under rotations, look like 
\begin{equation}
  S_{\f{\mu}\f{\nu}} = 
  \left(
  \begin{array}{cccc}
    S_{\f{0}\f{0}}(t,r) & S_{\f{0}\f{1}}(t,r) & 
      0                   & 0 \\
    S_{\f{0}\f{1}}(t,r) & S_{\f{1}\f{1}}(t,r) & 
      0                   & 0 \\
    0 & 0 & S_{\f{2}\f{2}}(t,r) & 0 \\
    0 & 0 & 0                   &  S_{\f{2}\f{2}}(t,r)
  \end{array}
  \right).
\end{equation}
This follows from the assumptions about the symmetry. 
\\
All components of $d_{abc}{}^d$ are either zero or proportional to
$d_{\f{1}\f{0}\f{1}}{}^{\f{0}}$.
\subsubsection{The resulting system}
Due to the complicated form of the matter terms the set of equations is
very lengthy. The equations are given in appendix \ref{AppsphSys}, they
are derived from system (\ref{quaSys}) and 
\begin{displaymath}
 \partial_r \left( \frac{- \N{e}^1{}_{{\f{0}}{\f{1}}}}{e_{\f{1}}{}^1}
\right) - \partial_t \left(
\frac{\N{e}^0{}_{{\f{0}}{\f{1}}}}{e_{\f{1}}{}^1} \right) = 0.
\end{displaymath} 
The system (\ref{quaSys}) has some remarkable features which should be
mentioned. Firstly it is semi linear. Secondly every equation has one of the
following forms:
\begin{eqnarray*}
  \partial_u {\cal U} & = & b_{\cal U}({\cal U},{\cal V},{\cal T}) \\
  \partial_v {\cal V} & = & b_{\cal V}({\cal U},{\cal V},{\cal T}) \\
  \partial_t  {\cal T} & = & b_{\cal T}({\cal U},{\cal V},{\cal T}), \\
\end{eqnarray*}
where ${\cal U}$, ${\cal V}$ and ${\cal T}$ are variables propagating
along $u$, $v$ and $t$. ${\cal U}$ stands for $\gamM{}$, $\hRi{}$ and
$\hphi{}$,  ${\cal V}$ for $\gamM{}$, $\hRiii{}$ and
$\hphiii{}$, ${\cal T}$ for $\ei{}$, $e$, $\gamii{}$, $\gamma{}$,
$\hRii{}$, $\Omega{}$, $\Omo{}$, $\Omi{}$, $\phi{}$, $\pho{}$,
$\phI{}$ and $\dioio{}$. For every ${\cal T}$ there also exists a
constraint,
\begin{equation*}
  \partial_r  {\cal T} = c_{\cal T}({\cal U},{\cal V},{\cal T}).
\end{equation*}
There are no constraints for the ${\cal U}$ and ${\cal V}$.
\subsubsection{The regularity conditions}
Polar coordinates cause regularity conditions in the center, in
physical spacetime as well as in unphysical spacetime. For the
variables in the system (\ref{sphSys}) those conditions are given in
appendix \ref{regularityConditions}. They express that locally the
center behaves like Minkowski space, and indicate how a wave hitting
the $r=0$ (inner) boundary is reflected there (passes through the
center). The regularity conditions at the inner boundary are the part of the
code that would change if a spacetime with a throat were
calculated. 
\\
But in unphysical spacetime there are also necessary 
conditions for the regularity at \Scri{} --- expressing an appropriate 
fall-off in physical spacetime and therefore reflecting asymptotical
flatness.
\\
It is not known whether those conditions are also sufficient for
regularity on \Scri{} in the general case. The answer to this requires
the investigation of the constraints on a given hypersurface. For the
case that a neighbourhood of \Scri{} is free of matter and for a special
choice of a hypersurface (for technical reasons) it has been shown
that the necessary conditions are also sufficient \cite{AnCA92ot}.
In the calculations presented here those conditions are fulfilled.
\\
There are as many regularity conditions (at the center and at \Scri{})
as there are variables ${\cal T}$. 
\section{The numerical method}
In this section the numerical methods used will be described shortly
and the reasons given for choosing those methods. 
\subsection{The initial value solver}
There are 12 constraints and 12 regularity conditions to be solved.
The free functions in the constraints are the six variables ${\cal U}$
and ${\cal V}$. But giving those makes the interpretation of a
parameter study difficult. ${\cal U}$ and ${\cal V}$ represent higher
derivatives of the primary quantities, the metric $g_{ab}$, the
rescaling factor $\Omega{}$ fixing the relation between physical and
unphysical spacetime and the scalar field $\phi{}$.
Furthermore the regularity conditions at \Scri{} are easier to handle
if it is known where $\Omega{}$ vanishes. A straightforward way to
realize this is to give $\Omega{}$ on the initial slice.
\\
In the code $e_{\f{1}}{}^1$, $\gamma^{\f{0}}{}_{\f{1}\f{1}}$,
$\gamma^{\f{0}}{}_{\f{2}\f{2}}$, $\Omega{}$, $\phi{}$, and
$\phi_{\f{0}}$ are given as free functions (because of regularity they
all must be even at the center). $e_{\f{1}}{}^1 =
e_{\f{0}}{}^0$ determines the $g_{00}$ respectively the $g_{11}$
component of the metric,  $\gamma^{\f{0}}{}_{\f{1}\f{1}}$ and 
$\gamma^{\f{0}}{}_{\f{2}\f{2}}$ the extrinsic curvature of the initial
slice in unphysical spacetime. The constraints which contain
derivatives of these quantities 
are interpreted as algebraic conditions for the quantities on the
right hand side. To ensure the invertibility of the resulting system,
including the points where $\Omega=0$, instead of equation
(\ref{ND1Om}) its derivative 
is used. A constraint for $\hRi+\hRiii{}$ results which becomes an
algebraic condition on $\Omega=0$ if $\hRi+\hRiii{}$ is at least
$C^1$ (condition \ref{smoothScri}). This is a necessary condition for
a regular \Scri{}. 
\\
The differential algebraic system has boundary conditions in the
center and at $\Omega=0$. As there are as many boundary conditions as
variables to solve for all the initial value freedom is coded in
the free functions 
$e_{\f{1}}{}^1$, $\gamma^{\f{0}}{}_{\f{1}\f{1}}$, 
$\gamma^{\f{0}}{}_{\f{2}\f{2}}$, $\Omega{}$, $\phi{}$, and
$\phi_{\f{0}}$. The system is solved with a relaxation scheme
combined with a Newton--Raphson solver derived with minor modifications
from the code given in \cite{PrFA88NR}. Firstly the system is solved
between $r=0$ and $\Omega=0$. If desired the same scheme can be used
to extend the initial hypersurface beyond the intersection with
\Scri{}. For this integration boundary conditions are given at
$\Omega=0$, namely the values obtained at $\Omega=0$ when solving the
constraints ``inside'' \Scri{}. To avoid a dependency of the values at
\Scri{} on the treatment of the outer grid boundary the initial
hypersurface has always be slightly extended beyond \Scri{}.
\\
To improve accuracy the system is solved with different grid sizes and
the results are Richardson (or Bulirsch--Stoer) \cite{PrFA88NR}
extrapolated to vanishing grid size. In the test cases, where an 
exact solution is known, the accuracy was limited by the rounding error
of the inversion of the matrix in the Newton--Raphson part of the code.
Typically the numerical and the exact solution differed in the last
two digits for calculations with 8 byte floating point numbers.
\subsection{The time integrator}
Constrainted evolution schemes are known to give in general more
accurate results than free evolution schemes. But there are 
disadvantages of constrainted evolution schemes. In a constrainted
evolution scheme the values at a certain grid point depend on the
values at many other grid points on the same time slice. Thus if one
grid point becomes singular the property of being singular is spread
over the
time slice into regions which are not causally connected with the
singular point.
\\
In most approaches used in numerical relativity the singularity is
avoided by an appropriate choice of the coordinate system, actually the
necessity of singularity avoidance has become a dogma of numerical
relativity \cite{BoM92av,ShT86rs}. A quite often used example for a
singularity avoiding coordinate system is obtained by the radial
gauge in spherically symmetric spacetimes \cite{Ch93ua,BaP83gr}. This
coordinate system cannot penetrate apparent 
horizons which are supposed to wrap around singularities by the cosmic
censorship hypothesis. The very interesting region of spacetime, where
the gravitational field has become strong enough to prevent light rays
from expanding, is excluded from being calculated. 
\\
The double nulllike coordinate system used does not avoid
singularities, the coordinate lines can only end at a singularity. In
this subsection it will be described how a program 
crash is avoided if singular points occur and how the calculation is
continued in the region of spacetime which are outside the future
domain of dependence of singular points. In the chosen approach the
singular boundary of spacetime is represented by grid points.
\subsubsection{Inside the grid}
To avoid the spreading of the singular property out of the domain of
influence of the continuous equations it is necessary to run
the scheme, at least in the very neighbourhood of a singularity, with a
Courant factor of $1$. 
Different schemes have been tested \cite{Hu93nu}. The only second
order schemes which could be run at a Courant factor of 1 were the
second order schemes of the class $S^\alpha_\beta{}$ \cite{PeT83CM}
with $\beta=1/2$. 
A second order Lax--Wendroff scheme as follows has been chosen: For the
equation $\partial_t f_{I} = \lambda_{I}  \partial_r f_{I}
+ b(\underline{f})$, with $\lambda_{I}=const$ and a vector of
functions $\underline{f}$, the predictor step is 
\begin{equation*}
  f_{\rm I}{}_{i-1/2}^{j+1/2} = 
    \frac{1}{2} \, \left( f_{\rm I}{}_{i-1}^j + f_{\rm I}{}_i^j \right)
    + \frac{\Delta t}{2 \Delta r} \, \lambda_{\rm I} \, 
        \left( f_{\rm I}{}_i^j - f_{\rm I}{}_{i-1}^j \right)
    + \frac{\Delta t}{2} \>
       b\left( \, \frac{1}{2} \left( \underline{f}\,{}_{i-1}^j +
                      \underline{f}\,{}_i^j \right) \, \right).
\end{equation*}
Then the corrector step is 
\begin{equation*}
  f_{\rm I}{}_i^{j+1} = 
    f_{\rm I}{}_i^j
    + \frac{\Delta t}{\Delta r} \, \lambda_{\rm I} \, 
        \left( f_{\rm I}{}_{i+1/2}^{j+1/2} - f_{\rm
          I}{}_{i-1/2}^{j+1/2} \right)
    + \Delta t \>
       b\left( \, \frac{1}{2} \left( \underline{f}\,{}_{i-1/2}^{j+1/2} +
                      \underline{f}\,{}_{i+1/2}^{j+1/2} \right) \, \right).
\end{equation*}
If $\lambda_{\rm I}$ were a function of $\underline{f}$, as it is the
case for other coordinate choices, it would be discretized as $b$.
\\
The second order Lax--Wendroff scheme is used to evolve gridpoints not
depending on the values at the boundary. 
\subsubsection{At the boundaries}
\label{BoundaryTreatment}
The treatment at the outer boundary is irrelevant as long as the
scheme remains stable for the following reason: The initial value
surface is extended beyond the intersection with \Scri{}. Running with
a Courant factor of $1$ the points inside and on \Scri{}, representing
the physical spacetime, do not even depend on the values at the outer
boundary which has no intersection with \Scri{}. Even if the treatment
of the outer boundary causes an instability it will 
not influence the physics, i.e.~the values on $\tilde M$ and \Scri{}.
In most runs with Coraunt factor $1$ the grid 
points depending on the outer boundary have not even been calculated.
For a run with a Courant factor $<1$ the influence of the outer
boundary treatment on the physical spacetime
must converge to $0$ with the same order the scheme
converges. Thus as long as the treatment on the outer boundary is
numerically stable the values at $\tilde M$ and \Scri{} are in the
limit of vanishing grid size independent of the outer boundary
treatment at any physical time. 
\\
Finding a stable treatment of the center was a very difficult problem.
So far I could not find a treatment which remains stable if a fourth order
scheme is built from the second order Lax--Wendroff by Richardson
extrapolation. 
\\
Especially it was not possible to extend the grid from the gridpoints
at $r=0$ and $r=\Delta r$ to $r=- \Delta r$ and run the Lax--Wendroff
scheme on the extended grid. 
\\
The solution of the stability problems at the inner boundary was to
replace the values near the center after every 
time step by the values obtained by other methods.
\\
In the method used for my Ph.D. thesis the constraints were integrated
from grid point number $2$ towards the center (the constraints together
with the regularity conditions determine all the variables at the
center). The
values at grid point $1$ have been obtained by interpolation. 
The solution did not look smooth on the
innermost gridpoints for coarse grids in the very strong field regime.
For very large fields (values of $A$ beyond $1.10$) this method even
became unstable. Therefore another method has been developed. 
\\
In the calculations for this paper a kind of polynomial extrapolation 
with dissipation has been used to replace the innermost gridpoints.
The idea is the following: Use the values at gridpoints $2$ and higher
and the regularity conditions to extrapolate towards the center by
polynomial 
fitting and get solution I. Do the polynomial fitting again starting
at grid point $3$ to get solution II. Solution I and II can now be
added in such a way that the simplest grid mode with values
$1,-1,1,\ldots{}$ is eliminated. This adding of dissipation is
necessary to ensure stability. I call this method polynomial
extrapolation with dissipation. 
\subsubsection{The singularity catcher}
\label{singCatcher}
Since the coordinate system does not avoid singularities some
variables may become singular. Depending on the default setting of the
compiler this causes a crash of the program by a floating point
exception (on UNIX systems the signal SIGFPE is sent). The programming
language C allows to 
specify what to do in case of a certain signal. In the program used
the action on SIGFPE is to flag the corresponding grid point as
singular and to continue the calculation on the rest of the grid.  
In addition to the reception of a SIGFPE signal a grid point is flagged
as singular if either the principal part, i.e.~$\ei{}$ or
$1-\frac{1}{4}\,\Omega^2\,\phi^2{}$, of the equation changes sign,
i.e.\ the
system of equations becomes singular in the sense of \cite{HuXXgr}, or
if the evaluation of the values
according to the scheme would involve points already flagged as
singular (hence called influence singularity). According to the latter
reason every point 
whose values depend on values at singular points must be flagged
singular. 
\\
A necessary condition for the stability of a scheme solving symmetric
hyperbolic systems is the Courant--Friedrichs--Lewy condition. The
domain of dependency of the discretized equations must be a superset
of the domain of dependency of the continuous equations, the Courant
factor must be $\le 1$. A Courant factor of $1$ means that both
domains of dependency coincide. On the other hand, every grid point
depending on singular gridpoints is singular. Thus all points in the
future null cone $L$ of a singular point must be flagged singular.  If
the Courant factor is $<1$ points outside $L$ also. For a scheme with
a Courant factor of exactly $1$ at least in the very vicinity of the
singularity it is possible to distinguish timelike/nulllike
singularities from spacelike singularities. 
If a line of gridpoints is numerically singular only because of a
influence singularity it is a strong hint for
a timelike singularity behind. Pure influence singularities appeared
only due to instabilities during the tests of different treatments of the
center. This is in agreement with cosmic censorship. 
\\
The dependence of the position of a singular line on the grid size has
been tested  intensely. In all calculated cases presented 
the position changed only by a negligible amount corresponding to a
few grid points for calculations with about 10000 spatial points. The
dependence of the first appearance of a singularity on the Courant
factor has also been tested. It only changed by a few grid points when
using Courant factors significantly smaller than $1$.
\section{Checking against exact solutions and accuracy estimates}
\subsection{Comparing with exact solutions}
On the exact solutions given below a number of tests have been
performed with
grids varying from $300$ to $10000$ spatial grid points and different
Courant factors. For most plots a coarser grid has been
used, typically with $100\times100$ grid points. 
\\
On the plotting grid the scheme shows the expected convergence
behaviour. For finer grids ($2000$ spatial points and
more) the error is dominated by second order terms. For coarser
grids fourth order terms (oscillations corresponding to high Fourier
modes) dominate the error near regions with steep gradients. 
\\
The error predictions by Richardson's extrapolation are in good
agreement with the actual error. 
\\
The violation of the constraints converges at least with second
order. Convergence is partly dominated by the error in the
discretization of the derivative for an
evaluation of the constraints. The violation of the constraints turned
out to be an inappropriate measure for the quality of the solution.
\subsubsection{Scalar field on flat background}
A scalar field on a flat background is obtained by setting $\kappa=0$
in the evolution equations. 
Since the physical energy-momentum tensor $\tT_{ab}$ is tracefree the
Ricci scalar $\tR{}$ in physical spacetime vanishes. $\tp{}$ is
determined by $\tilde{\vphantom{\phi}\Box} \tp = 0$. The corresponding
solution in unphysical spacetime is $\phi = \tp/\Omega{}$. On $\tilde
M$ the following rescaling has been used: 
\begin{equation}
\label{OmegaMink}
  \Omega(t,r) = 
  \frac{\pm 2}
       {\sqrt{(1 + \tan^2\frac{t+r}{2}) (1 + \tan^2\frac{t-r}{2}) }
         }
\end{equation}
with an obvious choice of the sign, depending on the position relative
to \Scri and $i+$. 
\begin{equation}
  \label{epeak}
  f(r) = 
    \left\{
    \begin{array}{c}
      e^{ -1/\left( 1-\sigma^2 (r-r_0)^2 \right) } \\
      0
    \end{array}
    \right. 
    \qquad \mbox{for} \qquad
    \begin{array}{c}
      \vphantom{e^{-1/\left( 1-\sigma^2 (r-r_0)^2 \right) } } 
        |r-r_0| < 1/\sigma \\
      \mbox{otherwise}
    \end{array}
    \quad ,
\end{equation}
with constants $\sigma{}$ and $r_0$. 
\\
This solution is useful for testing the matter part and the geometry
part of the code separately since $\kappa=0$ decouples their
evolution equations. But as most of the geometry variables are just
constants this is only a first step as test for the geometry part .
\subsubsection{By conformal gauge freedom deformed background}
For any positive function $\vartheta(t,r)$ the spacetimes
$(M,\vartheta^2 g_{ab},\phi/\vartheta)$ and $(M,g_{ab},\phi)$
correspond to the same physical 
spacetime if $(M,g_{ab},\phi)$ is a solution of the unphysical
equations. This gauge freedom can be used to obtain a unphysical
representation of Minkowski spacetime where the geometry variables
evolve in time and are not constant (except the Weyl tensor component
$\dioio{}$, which is zero for every representation of Minkowski space).   
For $\theta{}$ an ingoing wave pulse with the initial form
(\ref{epeak}) plus a constant $c$, such that $\theta\ge0$ everywhere,
has been chosen. $\phi{}$ has been set to $0$.  
\\
This solution has been used to make the geometry variables vary in
time and space and is thus a better test for the geometry part of the
code than the test in the previous subsection.
\subsubsection{$\phi+\const{}$ solution}
In physical spacetime $\tp=\const{}$ is a solution for Minkowski
spacetime. If we use $\const\ne0$ $\phi=\tp/\Omega{}$ is still a
solution on $\tilde M$ in unphysical spacetime. But this solution is
no longer regular on \Scri{} thus the calculation can be done in the
interior of $\tilde M$ only. For the tests that poses no
restriction, since in the unphysical spacetime there is nothing special
in the evolution equations on \Scri{}. 
\\
Since in this solution $\kappa\ne0$ and $\phi\ne0$  the coupling
between the geometry and the matter part can be tested.
\subsection{Error estimates and problems near the critical parameter}
\label{AccuracyProblem}
For the nonlinear regime two ways have been used to check the accuracy
of the solution. 
\\
The first is the Richardson
extrapolation to vanishing grid-size with the resulting error
estimates. The second uses the fact that the position of \Scri{} is
known and thus it is possible to compare the calculated value
$\Omega{}$ there with the exact value $0$. The error estimate for
$\Omega{}$ by the Richardson extrapolation was in all cases in
agreement with that deviation.
\\
On an approach to the critical parameter, the smallest parameter where a
singularity appears, there is a dramatic loss of accuracy although on
the coarse plotting grid the scheme still converges quadratically.
Viewing the fine grid used for the calculations one sees that the loss
of accuracy 
is largest around gridpoints 5--10, and this maxima of the error
decreases linear with the grid size, meaning that there is only
quadratic convergence in a $L^1$ norm, but not pointwise.  
\\
For models not to close to the critical value the accuracy problem
described can be cured by using more gridpoints. But with about 50000
spatial gridpoints that brute force method reaches its limitation.
Since the scheme converges quadratically in a $L^1$ norm only,
building a higher order scheme by Richardson extrapolation did not
work.
\\
A detailed inspection of the reasons hints, that this is partly a problem
of numerical regularization as described in \cite{Ev86aa}(pages
13--15). When the calculations for this paper were done there was no
way known to the author to numerically regularize the
Lax--Wendroff scheme. 
\section{Calculations}
In the calculations presented the free functions defining the
geometry of the initial slice are given as follows:
\begin{eqnarray}
  e_{\f{1}}{}^1 & = & 1 \\
  \gamma^{\f{0}}{}_{\f{1}\f{1}} & = & 0 \\
  \gamma^{\f{0}}{}_{\f{2}\f{2}} & = & 0 \\
  \Omega & = & 
       \frac{\pm 2}
       {\sqrt{(1 + \tan^2\frac{t_0+r}{2}) (1 + \tan^2\frac{t_0-r}{2}) }  }.
\end{eqnarray}
The value of the time coordinate on the initial slice $t_0$ is $\pi/2$.
The form of $\Omega{}$ is an often used choice for the rescaling of
Minkowski space. 
\\
For the scalar field the initial value is given by
\begin{equation}
\label{PolyPeak}
  \phi(r) = 
    \left\{
    \begin{array}{c}
      A \, \big( \quad 1 - 4 \, \sigma^2 (r-r_0)^2 + 6 \, \sigma^4 
(r-r_0)^4  \\
      \qquad \quad - 4 \, \sigma^6 (r-r_0)^6 + \sigma^8 (r-r_0)^8 
            \quad \big) \\ \\
      0
    \end{array}
    \right. 
    \qquad \mbox{for} \qquad
    \begin{array}{c}
      \vphantom{\begin{array}{c}\big( \\ \big)\end{array}}      
        |r-r_0| < 1/\sigma \\ \\
      \mbox{otherwise}
    \end{array}
    \quad ,
\end{equation} 
with $\sigma=(8/\pi)$ and $r_0 = \pi/4$. 
\\
This function $\phi(r)$ is the uniquely defined $C^3$-function with compact
support $[{r_0-1/\sigma},{r_0+1/\sigma}]$, being a polynomial of degree 8
on the support and with maximal value $A$ at $r_0$. $A$ is the parameter
to be varied. $\phi(r)$ should be at least $C^3$ in order to have all
variables in the system at least $C^1$ avoiding the known problems of a
Lax--Wendroff scheme at discontinuities. From the 
numerical viewpoint this form of the pulse is much better than a
$C^\infty{}$ partition of the one like (\ref{epeak}),  
since the variation of $\phi(r)$  and its spatial derivatives
is better distributed over the support.  
\\
$\phi_{\f{0}}$ is chosen in such a way that the pulse would be purely
ingoing on a flat background, which is the Einstein cylinder in
unphysical space, i.~e.
\begin{equation}
  \phi_{\f{0}} = \phi_{\f{1}} + \phi \, \frac{\cos r}{\sin r}
\end{equation}
for $r-r_0\in ]-1/\sigma,1/\sigma[$ and $0$ otherwise.
\\
A pulse with compact support provides several advantages when checking
the accuracy and interpreting the numerical solution since part of the
qualitative behaviour of the structure in the large is known from
analytic considerations. Figure \ref{globalStruct} shows the
qualitative behaviour in the large for a regular solution.
\insertfigure{7cm}{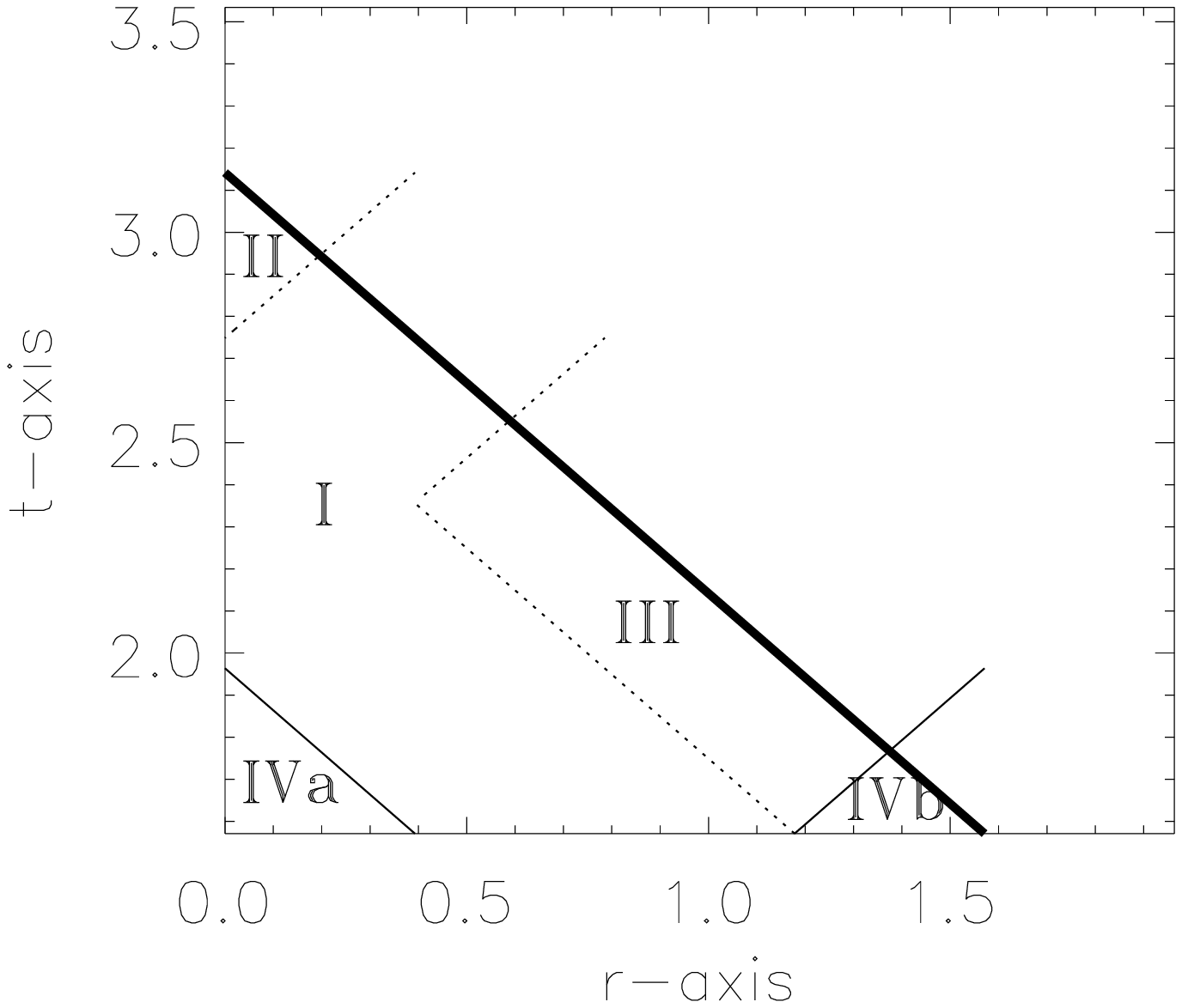}{\label{globalStruct}Qualitative
  picture of the structur in the large}
Region I is the compact support of the ingoing pulse in the linear
model ($\kappa=0$), passing through the
center and then crossing \Scri{}, the thick line. Every deviation from
a Schwarzschild 
solution in region III is caused by a pure back scattering effect. The
deviations from flat space in region II are also an back scattering
effect --- the field gets caught near the center for some time.  As
the characteristics 
have slope $1$, region IVa is a portion of flat space, region IVb is a
portion of Schwarzschild. For
stronger data the solution will become singular --- before reaching
the tip a singularity will appear. To investigate the structure
of this singularity is one of the goals of this work. In a regular
spacetime timelike infinity, $i^+$, lies at the point $(\pi,0)$.   
\subsection{A parameter study}
With increasing parameter $A$ more and more matter is put onto the initial
slice. At some critical parameter $A^*$ the spacetime is expected to
become singular. In this subsection the change of the global structure
will be discussed and a conjecture for the critical case presented. 
\\
When interpreting the behaviour with varying parameter two things have
to be kept in mind. 
\\
Firstly the characteristics, null geodesics, are a common
structure of all models. As the coordinates are adapted to this
structure it is straightforward and well defined to compare the
models.  
\\
Secondly the area distance $\tilde r :=
\sqrt{\tg_{\theta\theta}}$ to the center  in  physical spacetime,
expressed in unphysical variables, is 
\begin{equation*}
  \tilde r = \frac{r}{e \, \Omega}.
\end{equation*}
Even on the initial slice $e$ is determined as a solution of differential
equations, the constraints. Depending on $A$ the area distance of the
inner and outer edge of the shell of compact support of $\phi{}$ vary
(in this parameter study they monotonically grow with $A$). In
principle it is possible --- and for a better comparison with
M.~Choptuik's work desirable --- to give $e$ as a free function,
but parts of the initial value solver code would have to be
rewritten. But in doing so one can no longer give either $\Omega{}$ or
components of the extrinsic curvature as free functions. The first case
causes problems since the location of $S \cap \Scri{}$ on the
grid is no longer known, in the second case it is unknown whether the
necessary conditions for regularity on \Scri{} are also sufficient.  
\subsubsection{The conformal factor}
Figure \ref{Omega025} 
\insertfigure{7cm}{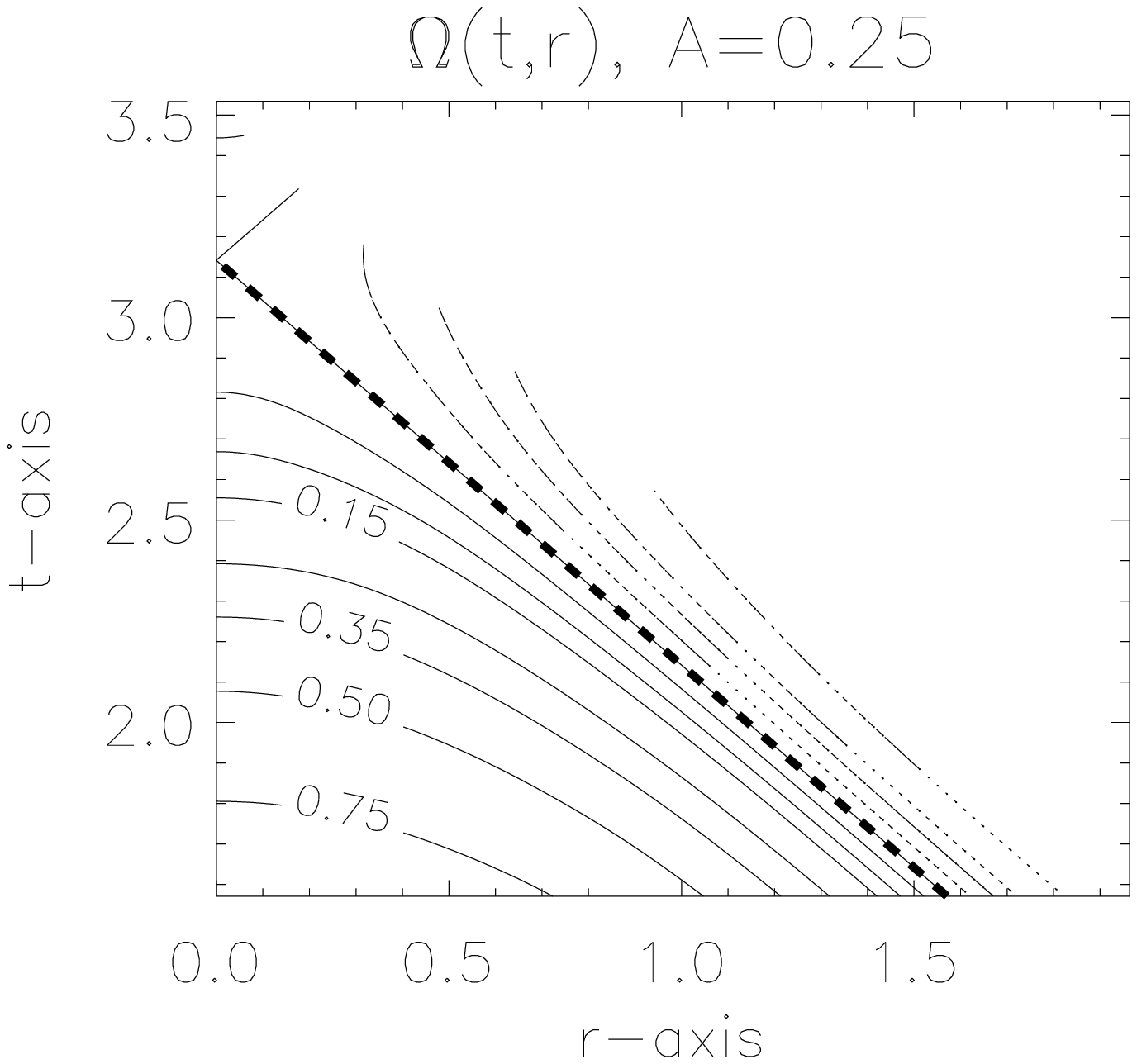}{\label{Omega025}$\Omega(t,r)$
  for $A=0.25$}
\insertfigure{7cm}{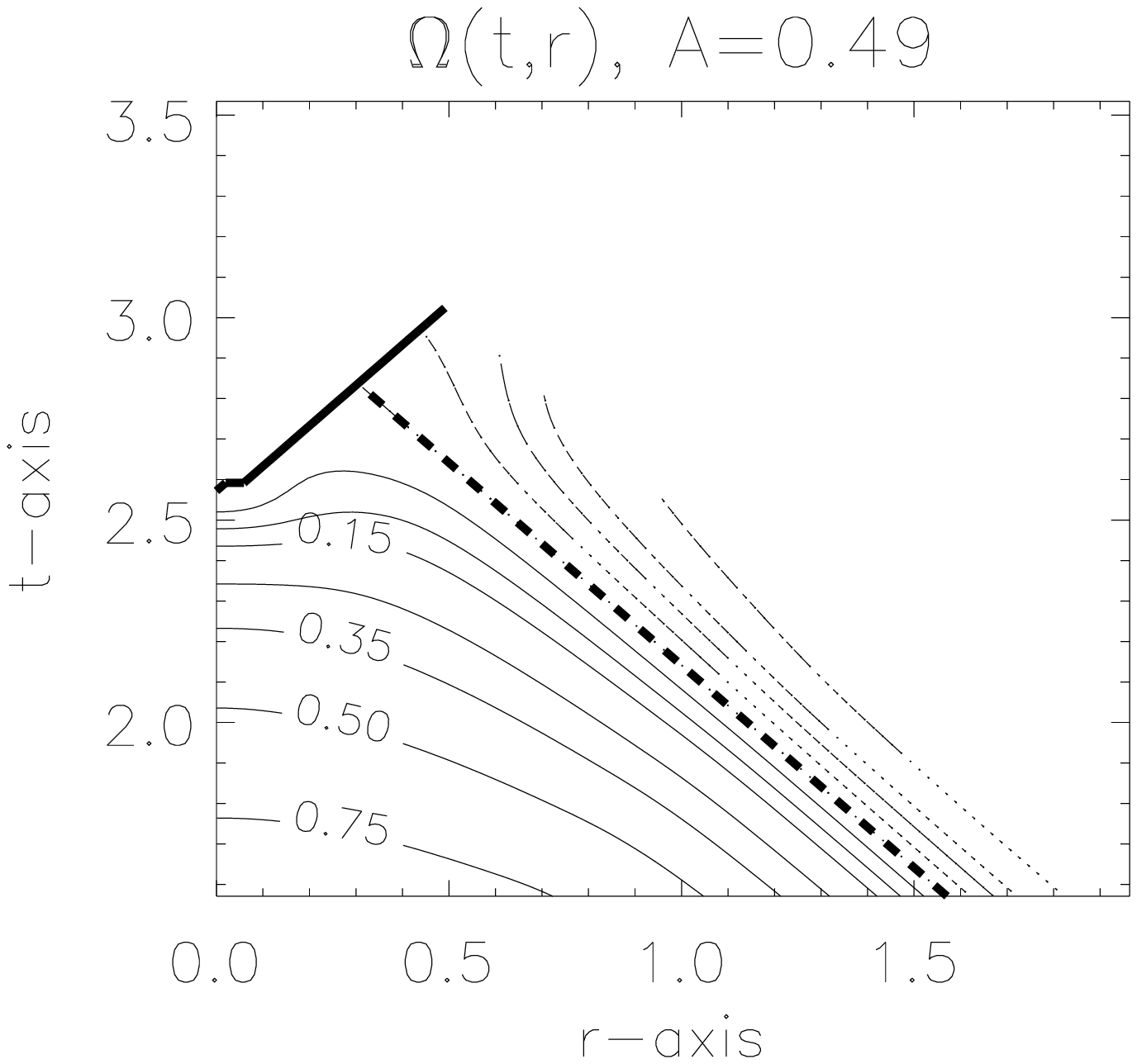}{\label{Omega049}$\Omega(t,r)$
  for $A=0.49$}
shows a contour plot of the rescaling factor $\Omega{}$ for
modestly strong initial data corresponding to a parameter $A=0.25$.
That is well below the critical value $A^*$, which is somewhere in the
interval $]0.48,0.49[$. Apart from some minor deformations in the
contour lines of $\Omega{}$, the figure looks
like Minkowski, although nonlinear effects like
backscattering are already significant. The $\Omega=0$ contour
coincides with 
the analytic location of $\Scri{}$ represented by the thick dashed
line. The values depending on the outer boundary of the grid have
not even been calculated. That is why there are no contours in the
upper triangle. 
\\
In figure \ref{Omega049} the spacetime has become singular.
The thick line is certainly a singular boundary of the future domain
of dependence of the initial data since here and in all the following
cases scalar invariants become singular. 
Near the center the singularity is spacelike. Further outside its
slope is numerically 
indistinguishable from a null line. The edgy look is an effect of the
plotting program and the coarse plotting grid, which will be
seen later when the
singularities will be examined more thorough --- a finer grid 
near the singularity showing more details will be used there. 
\\
With increasing amplitude of the initial scalar field (figure
\ref{Omega055}) 
\insertfigure{7cm}{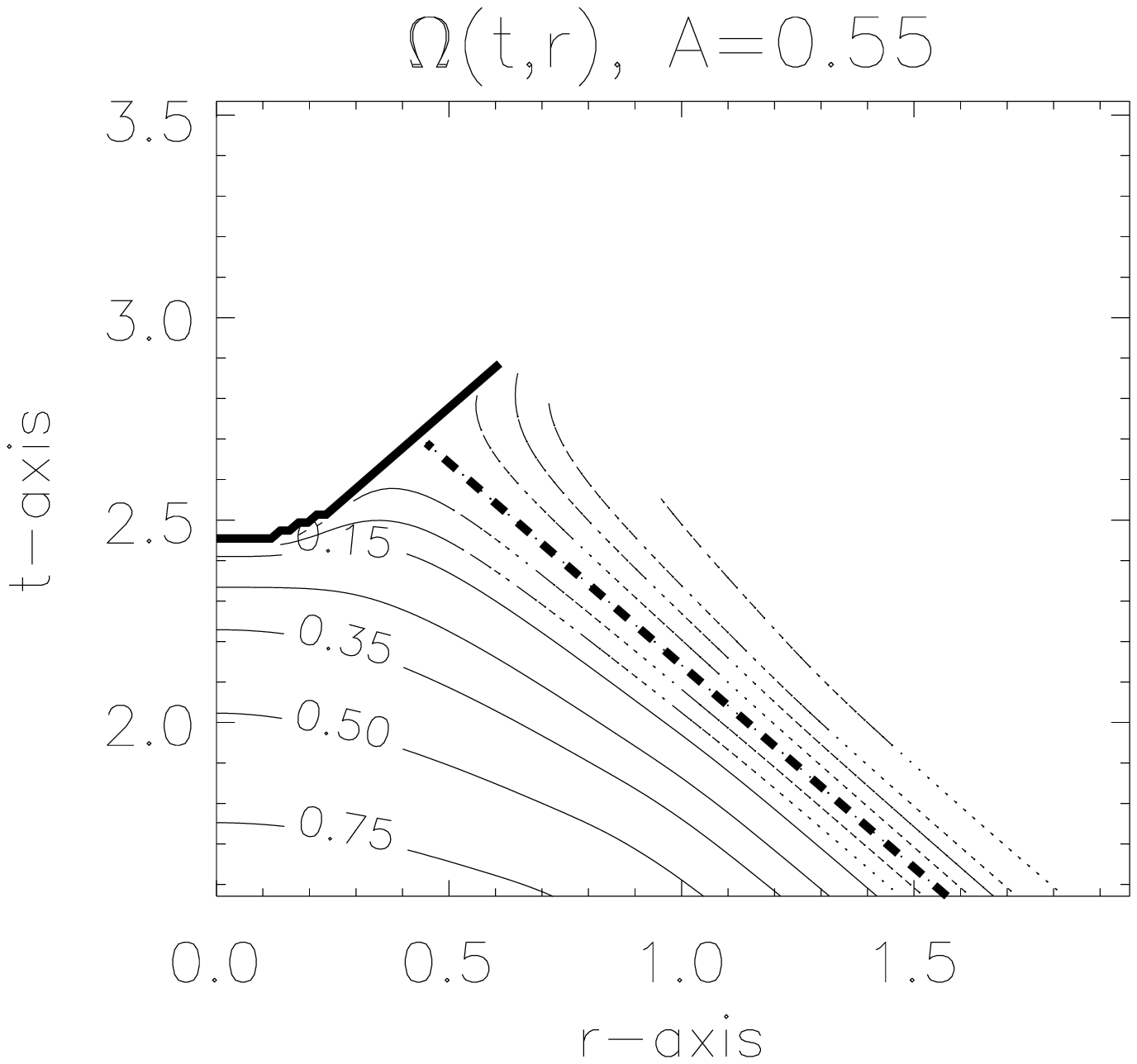}{\label{Omega055}$\Omega(t,r)$
  for $A=0.55$}
\insertfigure{7cm}{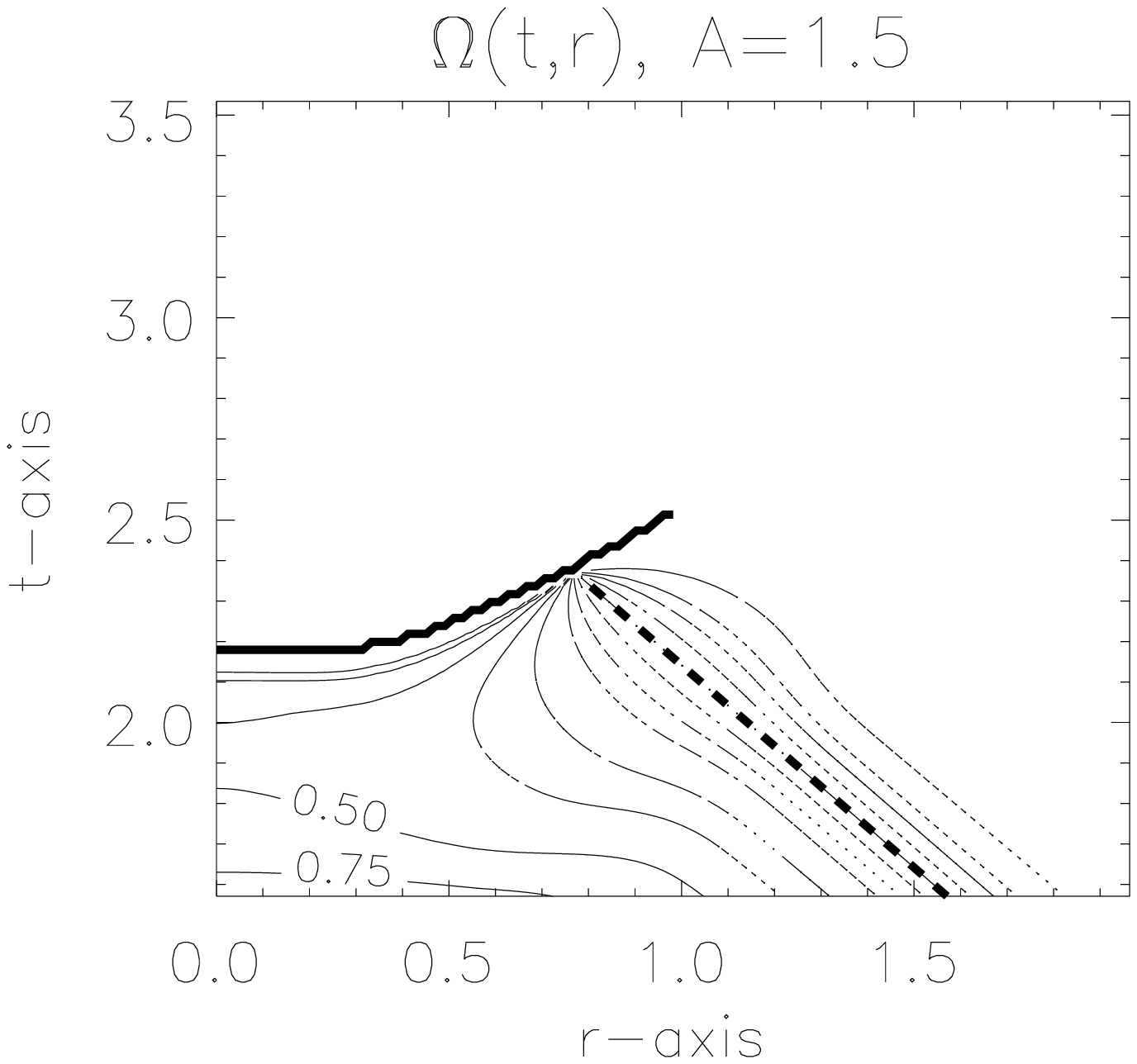}{\label{Omega150}$\Omega(t,r)$
  for $A=1.5$}
the spacelike part of the singularity has grown. The change of the
shape of the singularity on approaching the critical parameter from
above suggest the following conjecture: In the critical case the
singularity is a null line. 
\\
For further increasing $A$ the outer edge of the singularity becomes
spacelike. This part grows inwards and finally we get a picture like
figure \ref{Omega150}. Note the contour lines of $\Omega{}$ are
focused at the intersection of \Scri{} with the singularity.  Outside
\Scri{} the scalar $\Omega{}$ goes to $-\infty{}$, inside to
$\infty{}$ on approach to the singularity. In subsection
\ref{physicalReality} the structure of it will be examined. 
\\
The critical case can also be approached from the subcritical side. In
figure \ref{OmegaChange}
\insertfigure{10cm}{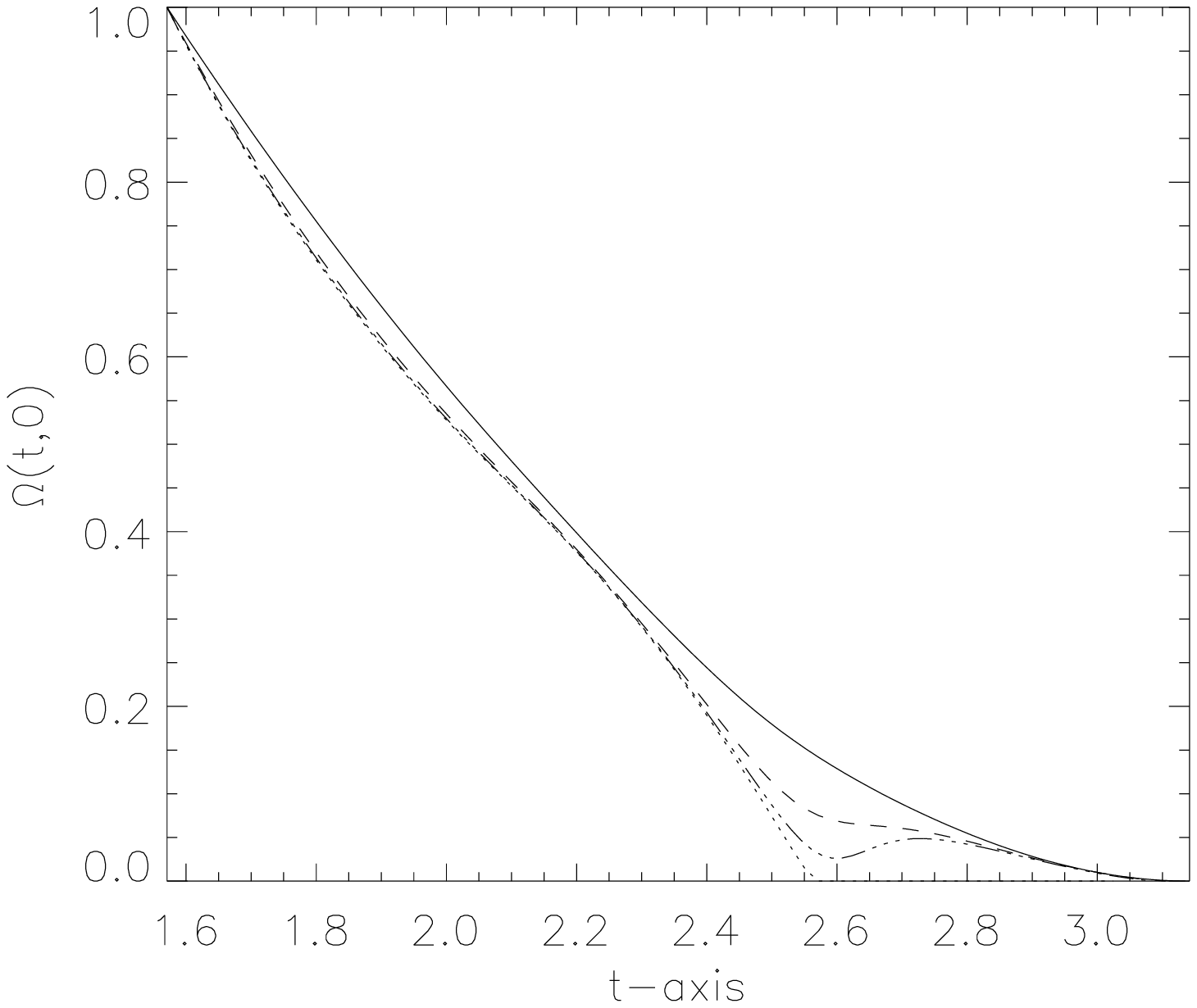}{
 \label{OmegaChange}$\Omega(t,0)$ for $A=0.25$ (---), $A=0.45$ ($--$),
 $A=0.48$ ($-\cdots-$) and $A=0.49$ ($\cdots$).}
the time dependence of $\Omega{}$ in the center is plotted for the three
subcritical values $A=0.25,0.45,0.48$ and the supercritical model
$A=0.49$.  
\\
Approaching the critical case $\Omega(t,0)$ seems to develop a zero
point before $i^+$. For spacetimes above the critical case the value
of $\Omega(t,0)$ at the singularity is greater than $0$ but again
approaching $0$ on approach to the critical case. Outside the center
and away from \Scri{}, the scalar $\Omega{}$ does not seem to go to $0$
even in the 
critical case. Although the values of $\Omega{}$ are influenced by
gauge, the statement that $\Omega{}$ vanishes is not. The author
cannot interpret this is exceptional conformal structure
for the critical case yet. Nevertheless the
calculations show the critical model might be recognized by this
behaviour of $\Omega{}$.
\subsubsection{Are the singularities found of physical nature?}
\label{physicalReality}
The figures of the preceding subsection show that there are
singularities for large values of $A$. In subsection \ref{singCatcher}
it has
been described when and how the program flags a point as singular. That does
not necessarily mean that those points represent a singularity in physical
spacetime. In this subsection arguments will be given to decide that
issue. Furthermore the quantities used classify the singularity. 
\\
In their singularity theorems R.~Penrose and S.~Hawking have shown
that a singularity is unavoidable if there are trapped surfaces and
the energy--momentum tensor fulfills certain energy conditions. The
conformally invariant scalar field does not fulfill those energy
condition. Nevertheless I regard the appearance of trapped surfaces as
a strong hint for singularities. With the transformation to the solution
of a massless Klein--Gordon field \cite{HuXXgr} it is easy to check
for the existence of an apparent horizon in a massless Klein--Gordon
field model which satisfies the relevant energy conditions. 
\\
The null expansion of in- and outgoing null directions in
spherical symmetry can be written as
\begin{equation}
\label{NullExpDef}
  \tilde\theta_{\I{out,in}} = \frac{1}{\tilde r} \, \left( \tilde
  e_{\I{u,v}}{}^a \tn_a \tilde r \right),
\end{equation}
where is the outgoing null vector and
\begin{displaymath}
  \tilde e_{\I{v}}{}^a = ( \tilde e_{\f{0}}{}^a - \tilde e_{\f{1}}{}^a
  )
\end{displaymath}
is the ingoing null vector. 
Written in unphysical quantities:
\begin{eqnarray}
\label{NullExpUnPhys}
  \tilde\theta_{\I{out}} & = & 
  \Omega \, \left(
  \gamma^{\f{0}}{}_{\f{2}\f{2}} - \frac{\gamma}{r} \right) -
  \Omega_{\f{0}} - \Omega_{\f{1}} 
\nonumber \\ 
  \tilde\theta_{\I{in}} & = & 
  \Omega \, \left( \gamma^{\f{0}}{}_{\f{2}\f{2}} +
  \frac{\gamma}{r} \right) - \Omega_{\f{0}} + \Omega_{\f{1}}
\end{eqnarray}
There is the freedom of null boost, thus
$\tilde\theta_{\I{out}}$ and $\tilde\theta_{\I{in}}$ are not gauge
invariant, but their product is. On the first view one might expect that
$\tilde\theta_{\I{out,in}}$ vanish for $\tilde r\rightarrow\infty{}$.
The gauge for $\tilde e_{\I{u,v}}$ chosen makes $\theta_{\I{out}}$ not
vanishing on $\Scri{}^+$, $\theta_{\I{in}}$ not vanishing on
$\Scri{}^-$. That is a pure gauge effect, by an appropriate null boost
$\theta_{\I{out}}$ and $\theta_{\I{in}}$ can be made vanishing on
\Scri{}.
\\
Another criteria for or against a ``real'' singularity is the
behaviour of curvature invariants of physical spacetime, e.g.~
\begin{equation*}
  \tilde C_{abcd} \, \tilde C^{abcd} = 
  \Omega^6 \, d_{abcd} \, d^{abcd} =
  12 \, \Omega^6 \, \left( d_{\f{1}\f{0}\f{1}}{}^{\f{0}} \right)^2.
\end{equation*}
Figure \ref{trapped055} 
\insertfigure{7cm}{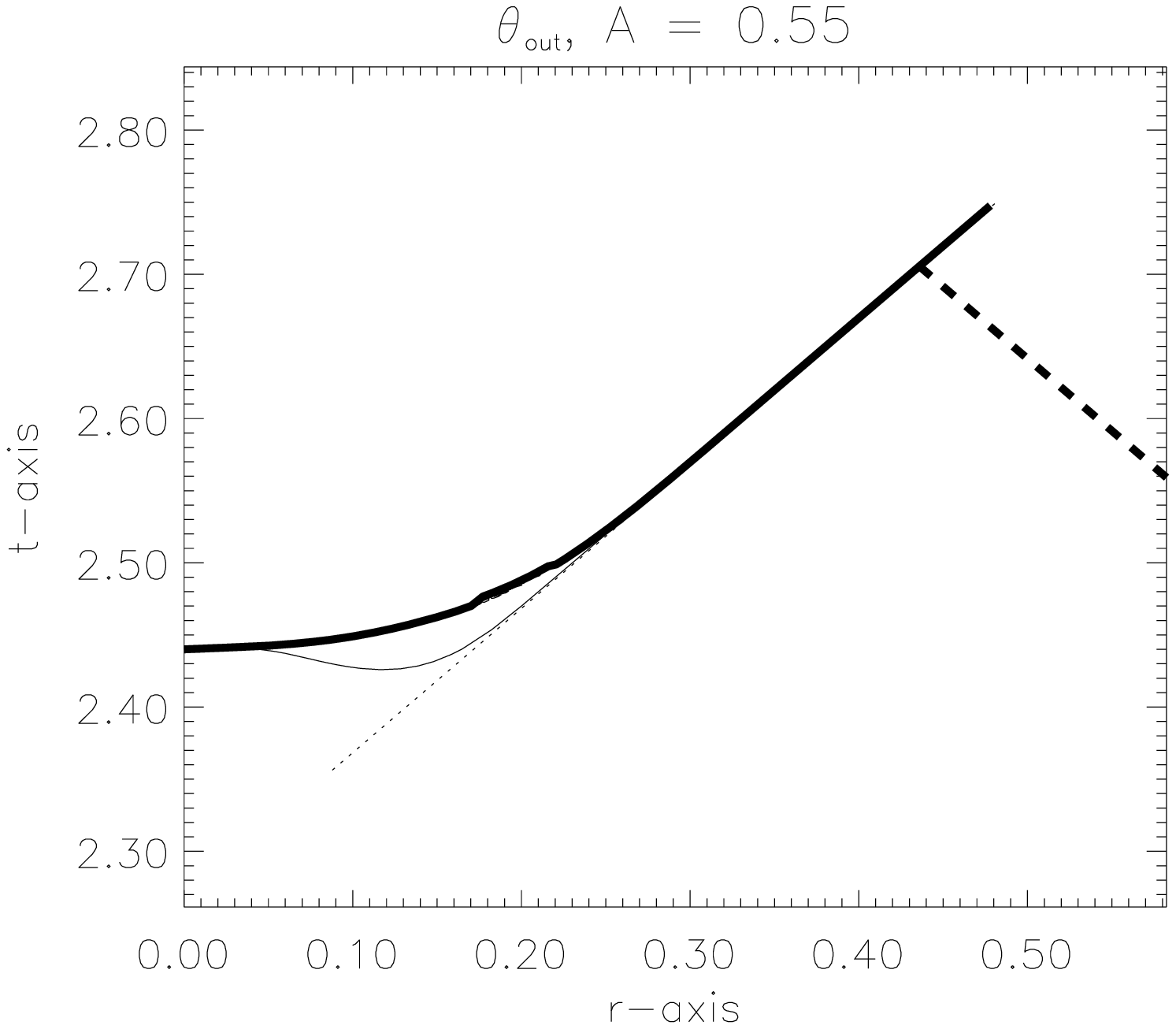}{\label{trapped055}$\theta_{\rm
    out}$ for $A=0.55$}
\insertfigure{7cm}{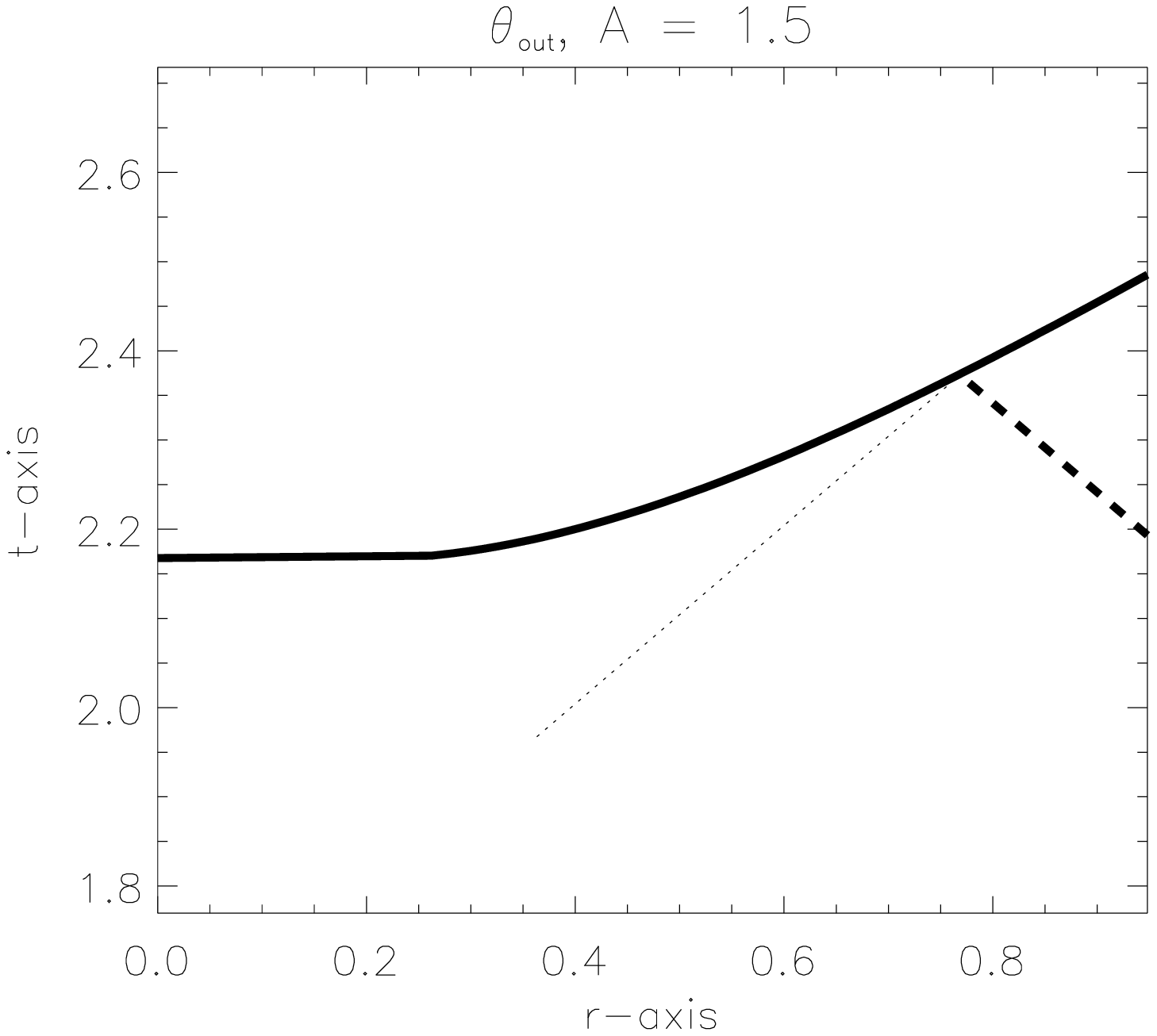}{\label{trapped150}$\theta_{\rm
    out}$ for $A=1.5$} 
shows the spacetime region near the
singularity of a $A=0.55$ model. The thick, dashed line is a part of
$\Scri{}^+$, the thick line the singularity ${\cal S}$. The
singularity is covered by a region of trapped surfaces, the thin line
shows the apparent horizon. In the corresponding initial value problem
for a massless Klein--Gordon field the same qualitative picture
arises.  
\\
The proper time of a Bondi observer (see equation
(\ref{BondiproperTime})) goes to infinity on approach to the
intersection point ${\cal P}$ of the 
${\cal S}$ with $\Scri^+$. This is a strong hint that ${\cal P}$ is a
singular timelike infinity, a singular $i^+$. Thus the last null line
from the center which reaches $\Scri^+$ at 
$i^+$ is the event horizon (dotted line).
\\
These are already strong hints that ${\cal S}$ is a singularity of the
physical spacetime. It is further strengthened by the fact that the
curvature scalar $\left(\tilde C_{abcd} \, \tilde
C^{abcd}\right)^{1/2}$ blows up
at least on approach to the inner (spacelike part) of ${\cal S}$.
There are also hints that the area radius goes to $0$ on ${\cal S}$,
but to decide that issue more numerical accuracy near the singularity
would be needed.
\\
All these considerations ``prove'' that ${\cal S}$ is a real
singularity of physical spacetime. 
\\
The situation is quite different in figure \ref{trapped150}. Along
${\cal S}$ the scalar $\Omega{}$ goes to $+\infty{}$ inside $\Scri{}^+$,
and to $-\infty{}$ outside $\Scri{}^+$. ${\cal S}$ is spacelike
everywhere. $\left(\tilde C_{abcd} \, \tilde C^{abcd}\right)^{1/2}$
does not show any sign of blow up near 
${\cal S}$, there is no region of trapped surfaces wrapped around
${\cal S}$. Due to the discontinuity of $\Omega{}$ near ${\cal P}$
the formula for the proper time of an observer at $\Scri{}^+$ is not
applicable. 
\\
Since only quantities of the unphysical spacetime become singular on
${\cal S}$ I conclude that ${\cal S}$ is a conformal
singularity caused by the gauge chosen by specifying $R(t,r)=6$. It is
very similar to a coordinate singularity. Only minor, unsuccessful
effort has been spent to try to avoid the conformal singularity by
using different choices of $R(t,r)$. 
\\
There is another very interesting point in the models with $A\ge1.2$.
On the initial slice there are regions where both
$\tilde\theta_{\I{in}}$ and $\tilde\theta_{\I{out}}$ are positive
(``antitrapped surfaces''). A calculation back in time shows a
singularity in the past, covered by an antitrapped region, i.e.~a
spacetime with a white hole. Although the model $A=1.2$ shows
a conformal singularity in the future the existence of an apparent
horizon suggests that the physical singularity is hidden behind the
conformal singularity. It is natural to assume that beyond $A^*$ there
is always a physical singularity which is sometimes hidden by a
conformal singularity.
\subsubsection{The mass scaling relation}
In his study of scalar fields M. Choptuik found the critical mass
scaling behaviour $m_{\rm BH} = a \left( A-A^*\right)^\gamma{}$ with $a$
independent of $A$ and an exponent $\gamma{}$ of approximately $0.37$.
\\
The Hawking mass
\begin{displaymath}
  m(t,r) = \frac{\tilde r}{2} \, \left( 1 + {\tilde r}^2 \,
  \tilde\theta_{\I{out}} \tilde\theta_{\I{in}} \right).
\end{displaymath}
can be written as 
\begin{eqnarray}
\label{Masse}
  m & = & \frac{1}{2} \, \left( \frac{r}{e} \right)^3 \,
  d_{\f{1}\f{0}\f{1}}{}^{\f{0}} \, ( 1 - \frac{1}{4} \kappa \Omega^2
  \phi^2 ) \nonumber \\ & & + \frac{1}{8} \, \left( \frac{r}{e}
\right)^3 \, \kappa \, \, \Omega \, \nonumber \\ & & \quad \left[
\phi^2 \, \left( \left( \gamma^{\f{0}}{}_{{\f{2}}{\f{2}}} \right)^2 -
\left( \frac{\gamma}{r} \right)^2 \right) + 2 \, \phi \, \left(
\gamma^{\f{0}}{}_{{\f{2}}{\f{2}}} \, \phi_{\f{0}} + \frac{\gamma}{r}
\phi_{\f{1}} \right) + \left( (\phi_{\f{0}})^2 - (\phi_{\f{1}})^2
\right) \right] \nonumber \\ & & + \frac{1}{8} \, \frac{r}{e} \,
\kappa \, \phi^2 \Omega.
\end{eqnarray}
The use of equation (\ref{SingGlSph}) on \Scri{} is essential
to write the 
Hawking mass in a form not containing terms with $\Omega^{-1}$ factors.
On \Scri{} the Hawking mass is identical with the Bondi mass. For
scalar field data with compact support on the initial slice in $\tilde
M$ the Bondi mass on the initial slice coincides with the ADM mass. 
\\
Figure \ref{scaling} 
\insertfigure{7cm}{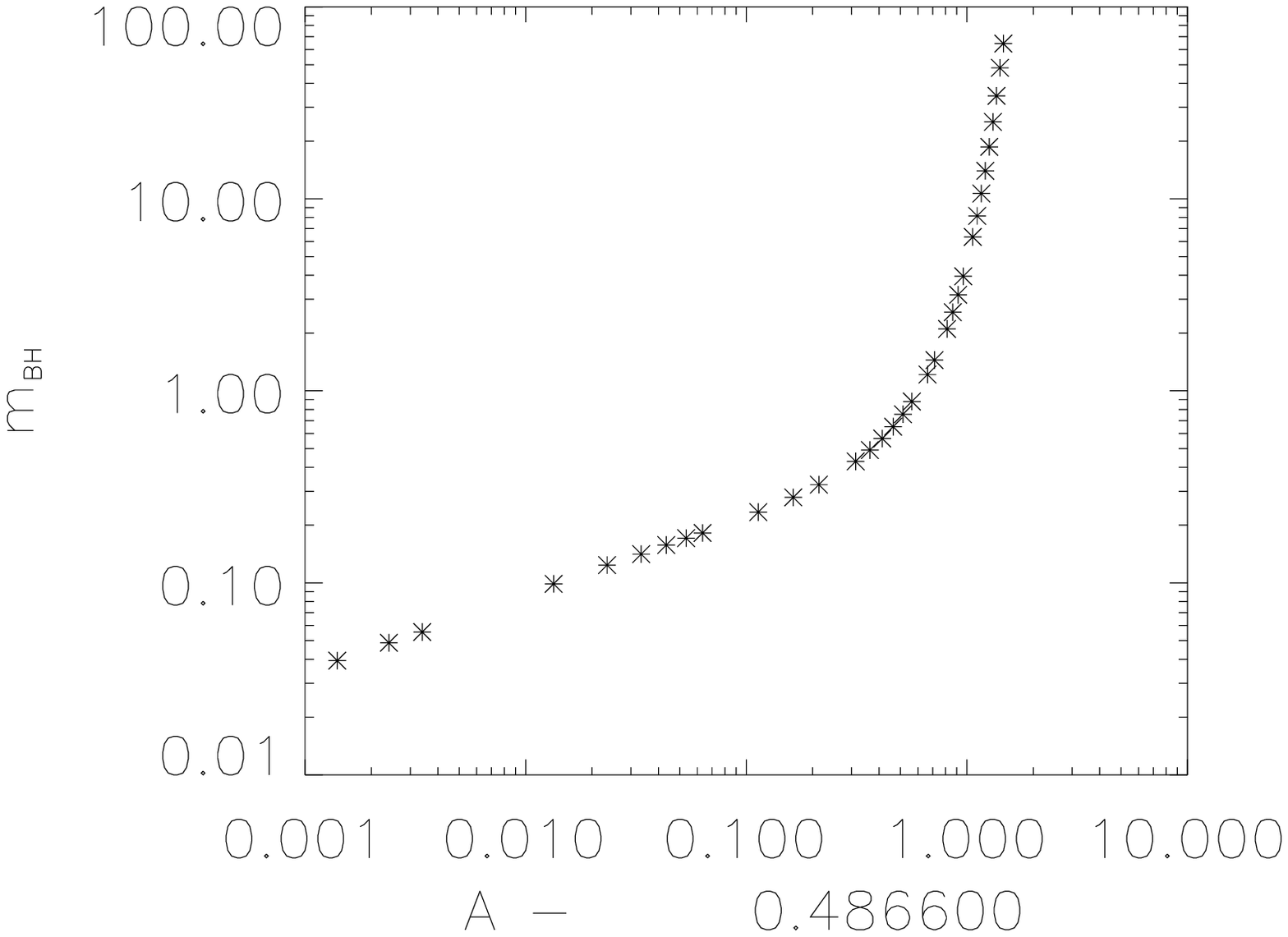}{\label{scaling}Scaling 
      relation for the black hole mass
      $m_{\rm BH}$}
\insertfigure{7cm}{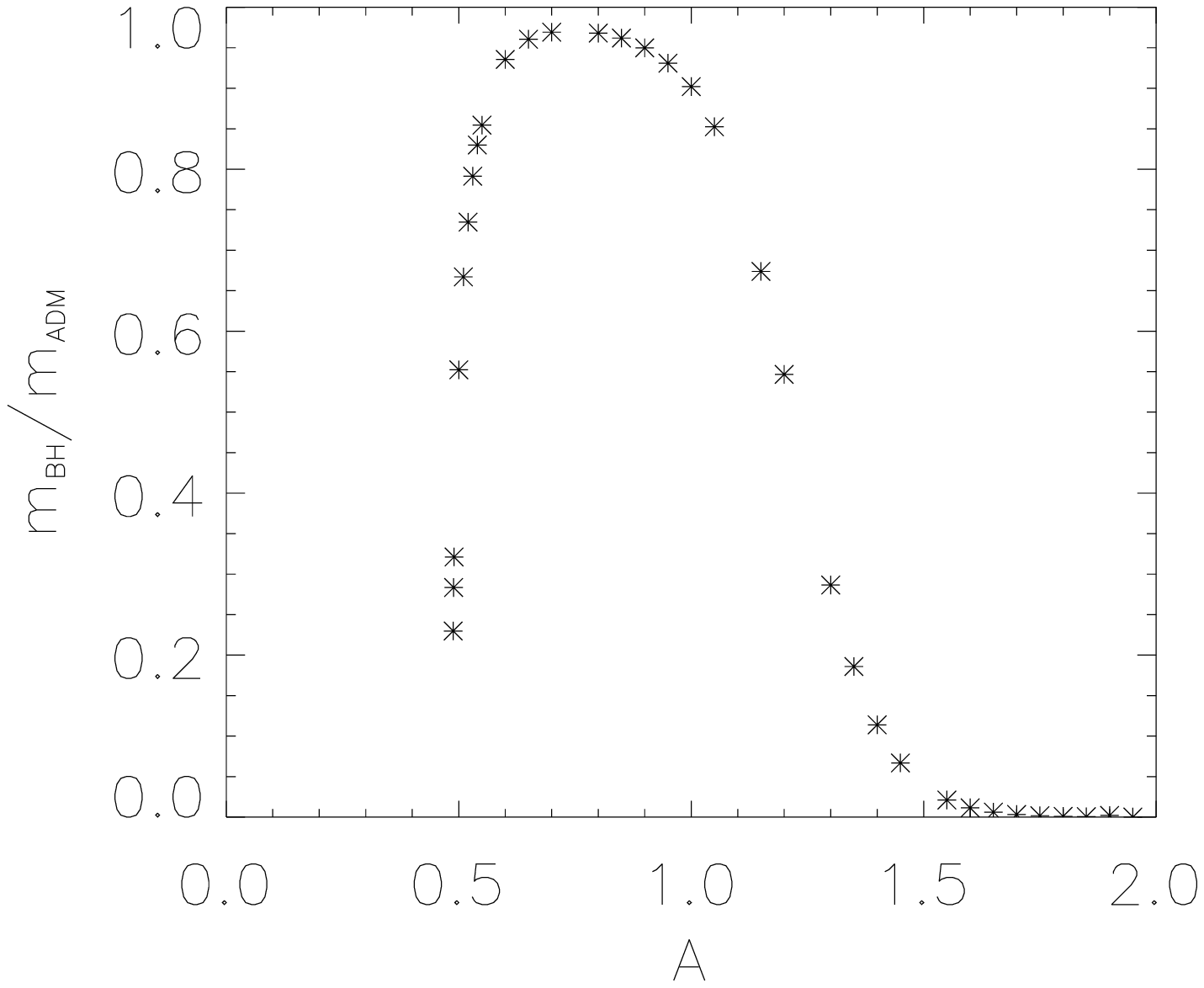}{\label{caught}Fraction of the 
      ADM mass $m_{\rm ADM}$
      caught by the black hole}
shows an logarithmic plot of the black hole mass obtained for the
parameter study performed here (the stars $\star{}$ mark calculated
models). The black hole mass is determined as the Bondi mass ``at
$i^+$''. 
\\
The value of $0.4866$ for the critical parameter $A^*$ has been found
by optimizing the straight line look in figure \ref{scaling}. 
\\
The critical exponent is $0.37$, but due to the uncertainty in $A^*$
the error is certainly $\pm 0.05$. Hence the results are compatible
with the results of others, but I would not claim more. 
\\
Although the mass caught by the black hole increases the percentage of
the caught mass has a significant maximum at $A\approx 0.75$. For
large $A$ almost all the mass escapes to $\Scri{}^+$. Actually
$A>\approx1.0$ the given value is only an upper bound since $i^+$ is
hidden behind the conformal singularity and thus the Bondi mass at the
intersection of $\Scri{}^+$ with the conformal singularity may further
decrease. The figures for the black hole mass of the corresponding
Klein--Gordon models show the same behaviour.
\\
For models with large initial scalar field amplitudes $A$ there are
number of significant changes in the 
structure of spacetime. Although most of the scalar field is still
contained in region I the shell with most of the Hawking
mass in it expands and crosses \Scri{} in region III. Mass and scalar
field ``decouple'', propagation of mass is almost completely
determined by nonlinear terms.   
\subsection{Christodoulou's theorem}
D.~Christodoulou investigated the general relativistic massless
Klein--Gordon field in spherically symmetric spacetimes in a number of
papers, for a list of references see \cite{Ch91tf}. He found
sufficient conditions for the appearance of a spacetime singularity.
The theorems proven there contain the following statement:
\begin{Theorem}
\label{ChrisSatz}\cite{Ch91tf}
Consider on the initial future null geodesic cone $\bar C_0^+$ an
annular region bounded by the two spheres $\bar{\Bbb{S}}_{1,0}$ and
$\bar{\Bbb{S}}_{2,0}$ with $\bar{\Bbb{S}}_{2,0}$ in the exterior of
$\bar{\Bbb{S}}_{1,0}$ and areal radii $\bar r_{1,0}$ and $\bar
r_{2,0}$. Let 
$$
  \delta_0 := \frac{\bar r_{2,0}}{\bar r_{1,0}} - 1 \quad \in
              \Big(0,\frac{1}{2}\Big),
$$
$$
  \eta_0 = \frac{2\,\left(\bar m_{2,0}-\bar m_{1,0}\right)}{\bar r_{2,0}},
$$
and
$$ 
  E(y) := \frac{y}{(1+y)^2} \, 
          \left[ \log\left(\frac{1}{2y}\right) + 5 - y \right].
$$
A sufficient condition for a non--timelike singular boundary in the
future of $\bar
C_0^+$ is 
$$
  \eta_0 \ge E(\delta_0).
$$
\end{Theorem}
The model with a conformal scalar field calculated here is (almost)
equivalent to the massless Klein--Gordon field \cite{HuXXgr}. If
$(\tilde M,\tg_{ab},\tp)$ is a solution for the general relativistic
conformal scalar field model $(\bar M,\bar g_{ab},\bar \phi)$ is a
solution for the general relativistic massless Klein--Gordon field with
\begin{eqnarray}
  {\B{\phi}}    & = &       
    \sqrt{6} \, \mbox{arctanh} \frac{\tp}{2}
  \\
  {\B{g}}_{ab}  & = & ( 1-\frac{1}{4}\tp^2 ) \, \tg_{ab}.
\end{eqnarray}
The area radius $\bar r$ and the Hawking mass $\bar m$ are given by 
\begin{eqnarray}
  \bar r & = & \omega \, \tilde r \nonumber 
\\ 
  \bar m & = & \omega \, m \, +
  \, \frac{{\tilde r}^3}{2} \, \left( \tilde \theta_{\I{out}} \,
  \tilde e_{\I{v}}(\omega) + \tilde \theta_{\I{in}} \, \tilde
  e_{\I{u}}(\omega) + \frac{1}{\omega} \, \tilde e_{\I{v}}(\omega) \,
  \tilde e_{\I{u}}(\omega) \right),  \nonumber
\end{eqnarray}
$\omega = \sqrt{1-\frac{1}{4}\Omega^2\phi^2}$. $\tilde r$ is the area
radius, $m$ the Hawking mass,  $\tilde \theta_{\I{out,in}}$ the null
expansion of the out- and ingoing null directions with null vectors
$\tilde e_{\I{u}}$ and  $\tilde e_{\I{u}}$ in the conformal scalar
field model. $\bar m$ can be written in a form which does not, even
implicitly, contain terms proportional to $\Omega^{-1}$. For the
purpose of this subsection it is not necessary to write $\bar m$ in a
form obviously regular on \Scri{}.
\\
In figure \ref{ChrisInitial} 
\insertfigure{10cm}{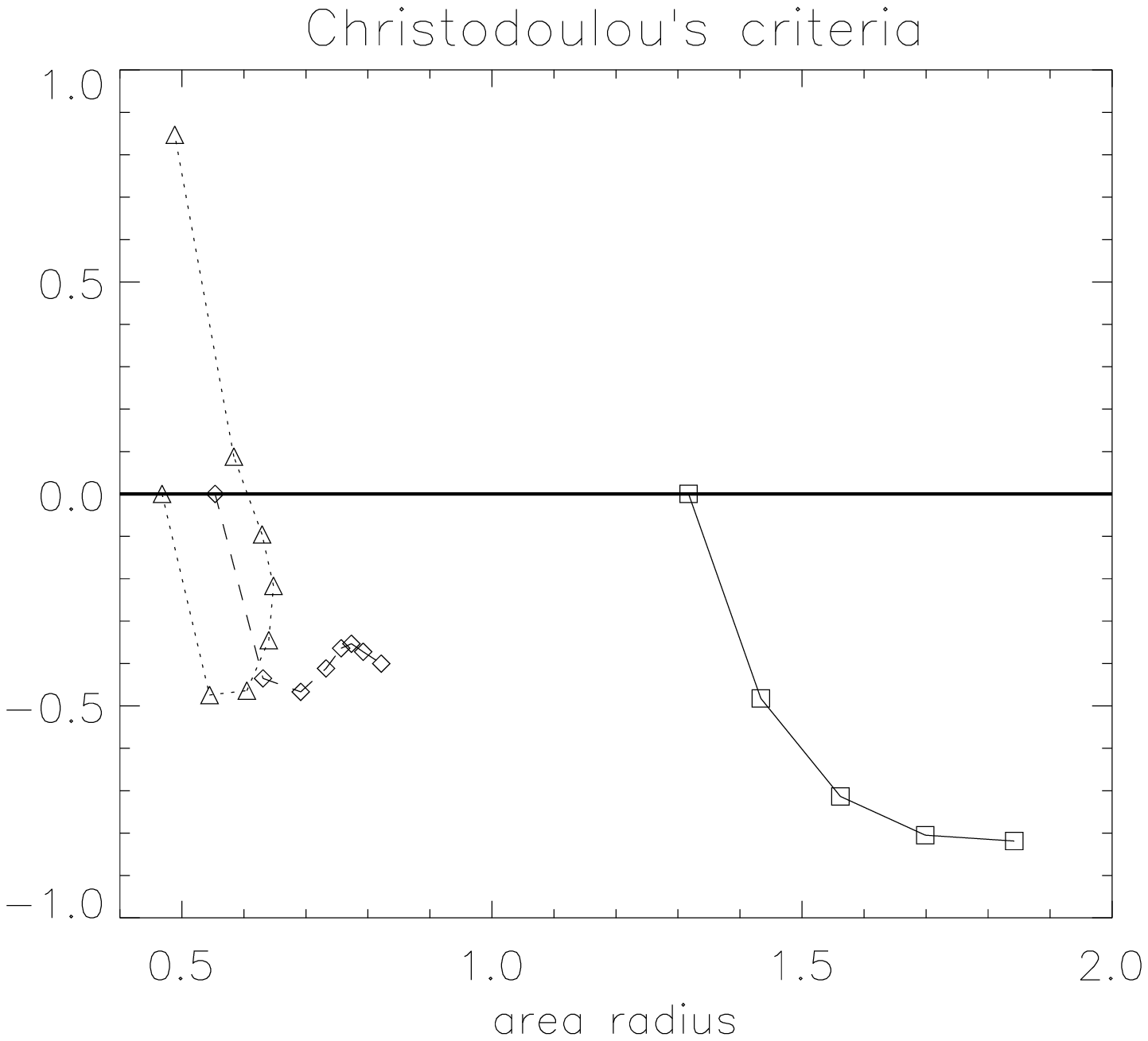}{\label{ChrisInitial}
      Christodoulou's criteria $\eta_0-
      E(\delta_0)$ versus area radius $\bar r$ for various 
      light cones}
%
$\eta_0 - E(\delta_0)$ is plotted for three outgoing light cones versus
the area radius $\bar r$. For
a parameter of $A=0.75$, a model far beyond the critical value with a
distinct singularity in physical spacetime, the solution has been
printed out for 100  
time slices with 100 grid points. On outgoing null cones $\eta_0
- E(\delta_0)$ has been evaluated. The sufficient condition is
satisfied if $\eta_0- E(\delta_0)\ge 0$. The gridpoints are marked with
squares, diamonds, and triangles. The first light cone (squares) lies
outside the event horizon, the criteria is nowhere fulfilled. The
second lightcone (diamonds) approximately coincides with the event
horizon. The criteria is not fulfilled either. Only if the
apparent horizon is crossed (triangles) $\eta_0- E(\delta_0)$ becomes
larger than 
$0$. Although many outgoing lightcones in various
singular models have been checked the criteria was only fulfilled
when the lightcone 
crossed the apparent horizon. Thus I conclude that the criteria is not
very sharp. This is in agreement with the results in \cite{GoW92na}.
\subsection{Extraction of Radiation}
In addition to the already presented determination of the Bondi mass I
am going to demonstrate the simplicity of
radiation extraction on two more examples in this subsection. In the
first case the effects of 
the nonlinearities in the equations are analyzed --- the purely
backscattered radiation in region III and  the decay of the radiation
in region IV of a regular spacetime model. In the second case values
of the Bondi mass in a singular model are compared with values of the
mass read off at finite radii. 
\subsubsection{Effects of non--linearity}
Figure \ref{radall} 
\insertfigure{7cm}{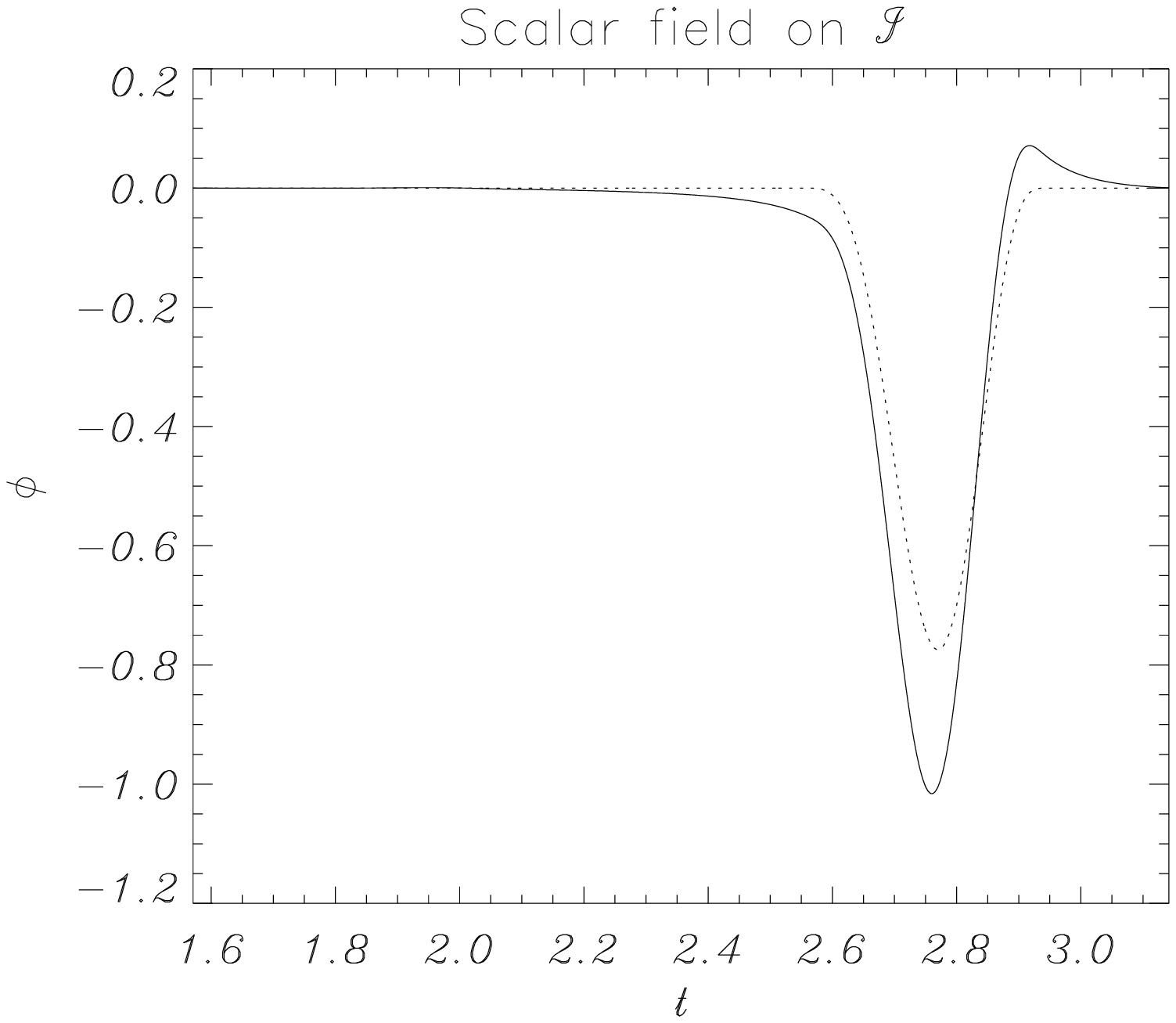}{\label{radall}$\phi{}$ for
  the whole time evolution}
%
compares $\phi{}$ on \Scri{}, i.e.\ the
coefficient of the $1/\tilde r$ term of the scalar field $\tp{}$ in physical
spacetime, for a model with an initial amplitude of $A=0.40$ for the
linear model 
($\kappa=0$, dotted line) and the nonlinear model ($\kappa=1$,
continuous line). In the model with gravitation a significant amount of
radiation, reflected outward by backscattering, has already crossed
\Scri{} before the linear signal reaches \Scri{} (region
III). The main signal is stronger in the nonlinear case and there 
is some scalar field left in region II.
\\
In the figure \ref{pureBackscatter} 
\insertfigure{7cm}{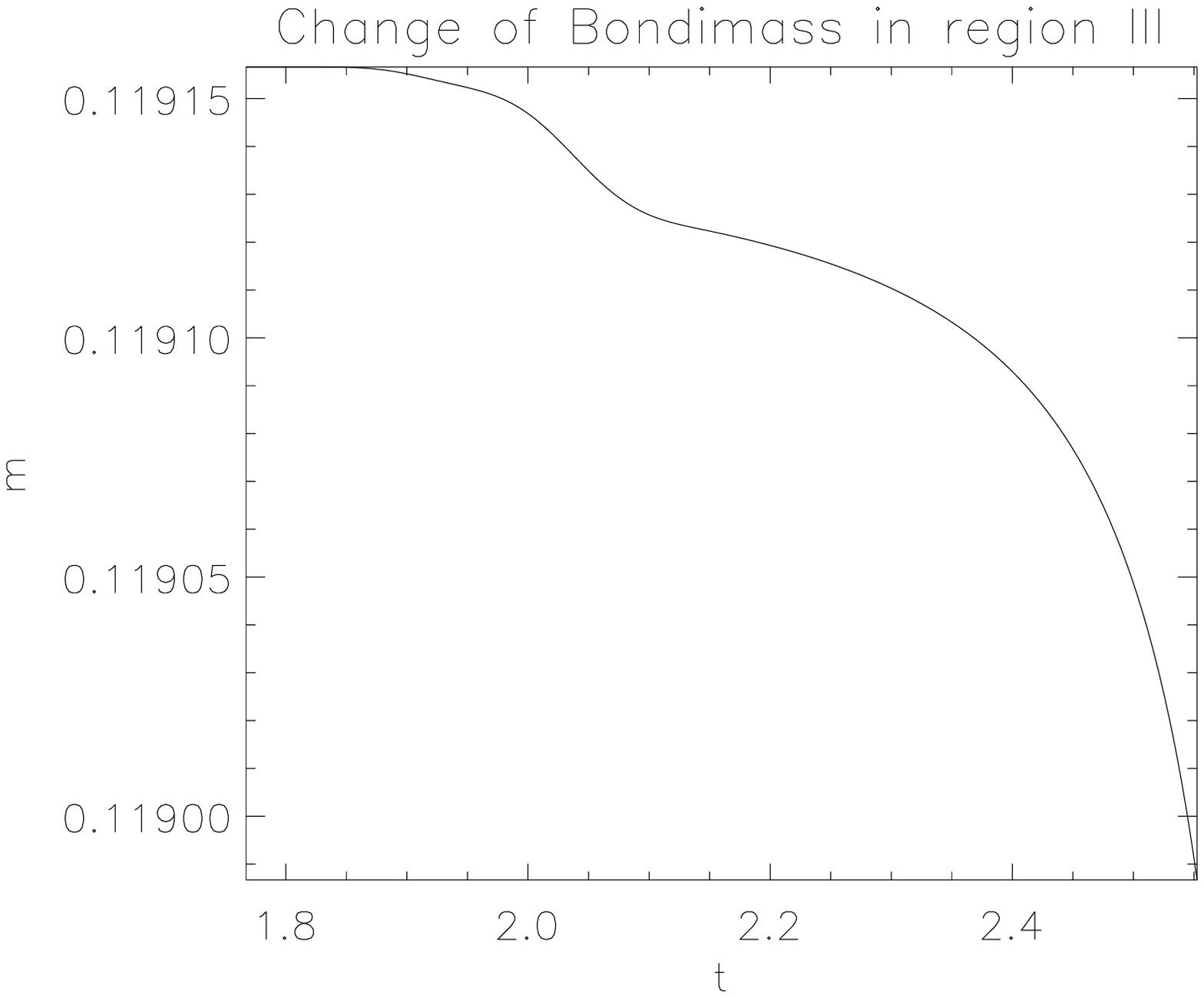}{
    \label{pureBackscatter}$m_{\rm
    Bondi}$ in region III} 
\insertfigure{7cm}{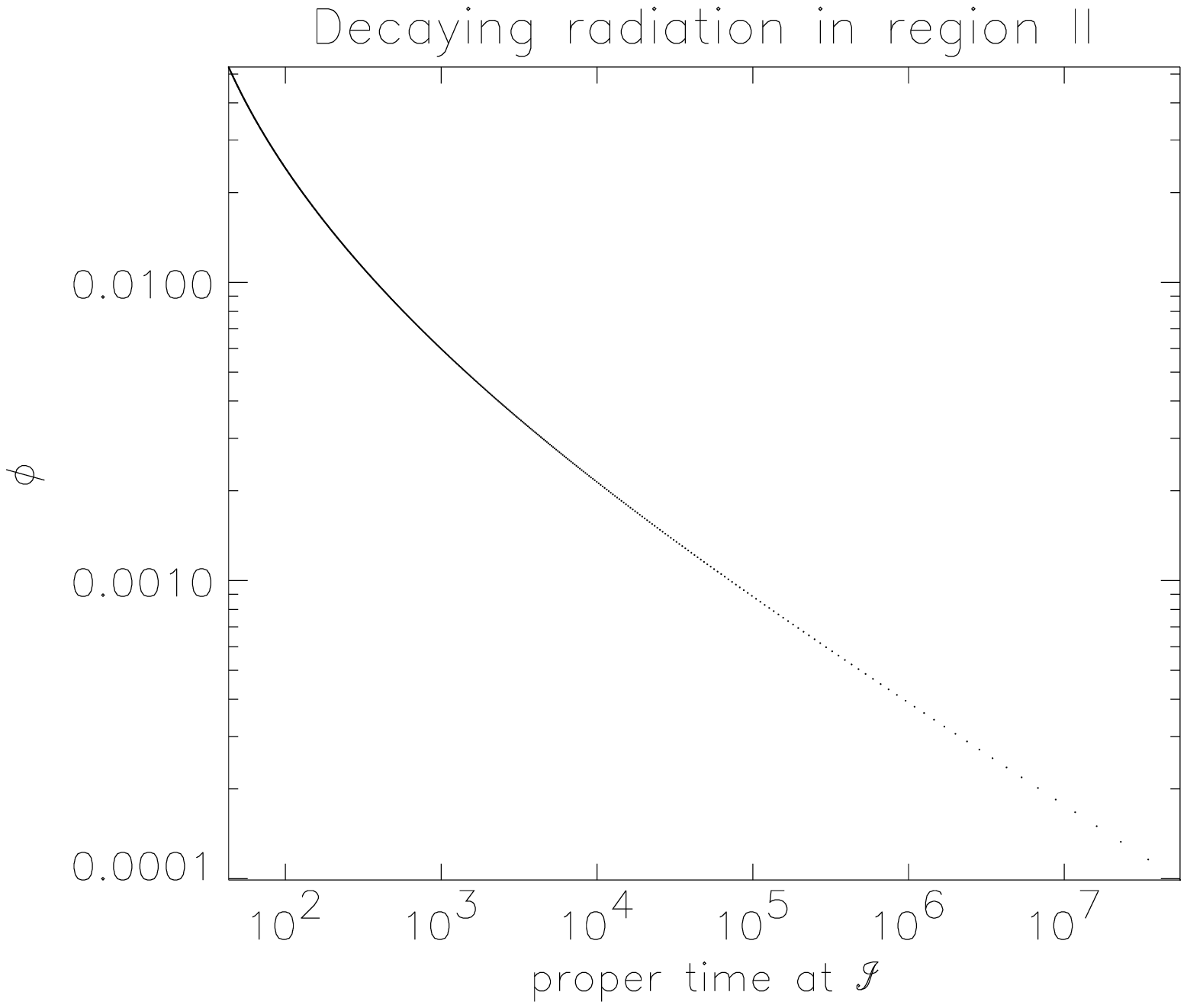}{
    \label{pureRest}$\phi{}$ in region II} 
%
the Bondi mass is shown in the pure
backscatter region III. Later times correspond to the backscattering
of a matter shell which has decreased in size during the infall. 
\\
The scalar field $\phi{}$ is displayed in a double logarithmic plot of
the scalar
field versus the proper time $\tau{}$ of an observer at \Scri{}. 
This observer
is obtained by taking a world line at fixed area radius $\tilde r$ 
(and angle)
and taking the limit $\tilde r \rightarrow \infty{}$. In the
unphysical variables one gets
\begin{equation}
 \label{BondiproperTime}
  \tau-\tau_0 = \int_{t_0}^t a\left(r_{\Scri}(t)\right) \, dt,
\end{equation}
where $r_{\Scri}(t)$ is the coordinate $r$ in unphysical spacetime of
\Scri{} at unphysical time $t$ and
\begin{eqnarray}
\label{BondimetricKoeff}
a & = & - \, \frac{1}{(\Omi)^2} \,
      \Bigg[  \, \frac{1}{(\ei)^2}  \,                           
          \left(  \frac{ - R - 6\, \hRii }{12} + (\gamii)^2       
          - \frac{1}{{r_{\Scri}}^2} \, ( \ei + \gamma )^2 \right) 
\\ \nonumber
  &     &  \hphantom{  - \frac{1}{(\Omi)^2} \Bigg[  } \quad      
          + \frac{1}{{r_{\Scri}}^2 \, \ei} \, ( \ei + \gamma )      
          - \frac{1}{{r_{\Scri}}^2}  \, \Bigg].
\end{eqnarray}
The dots represent grid points of the calculation with 5000 grid
points. Since on approach to $i^+$ the center is approached and the
values become inaccurate, only those points have been included in the
plot where the results of a Richardson extrapolation with 5000 and
10000 grid points are visually indistinguishable from those of an
extrapolation with 10000 and 20000 grid points. In this plot
the last 10 (of 5000) points on \Scri{} before $i^+$ are missing. 
\\
The slope in the asymptotical region is approximately $0.34$. Thus for
very late times $\phi \sim (\tau-\tau_0)^{0.34}$.
\\
Unfortunately the Bondi mass falls off so fast 
that rounding errors of the graphic program, with which the
calculation of the mass is done and which has only single precision,
do not allow to enter the asymptotical region with the constant slope
for $\phi{}$.   
\\
But in spite of the accuracy problem near the center, figure
\ref{pureRest} illustrates an advantage of the conformal method: The
time scale of the collapse of the initial matter shell is of order
$1$, the fall off of the decay could be investigated for almost $10^8$
times the dynamical time scale. 
\subsubsection{Radiative quantities at finite physical distance and at
  \Scri{}}
A rough estimate of the errors which are to be expected by reading of
radiative quantities at finite radius has been made in the following
way: Given a lower bound for the area radius, the grid on a time slice
has been 
searched for the first grid point $P$ with larger area radius. At that
point $P$ the
value for the Hawking mass $m_{\rm Haw, fin}$ and the ``finite'' Bondi mass,
\begin{equation*}
  m_{\rm Bondi, fin}(P) := \left. \frac{1}{2} \, \left( \frac{r}{e}
\right)^3 \, d_{\f{1}\f{0}\f{1}}{}^{\f{0}} \right|_P ,
\end{equation*}
have been calculated. Those values have been compared with the Bondi
mass on the intersection point of the future directed light cone of
$P$ and $\Scri{}^+$. The model used corresponds to the parameter $A=0.55$
and develops strong gravitational fields,
the spacetime has a singularity which catches approximately $85\%$ of
the ADM mass of $0.22$.  The outer boundary of the compact support of
the scalar
field lies at an area radius $\tilde r=2.7$, $P$ has been chosen two, five
and ten times that value, expressed in units of the
ADM mass, $P$ is at $10$, $25$ or $50$.
\\
The maximal relative error, i.~e.~error over ADM mass, is
approximately $2\%$, $1\%$ and $0.1\%$.
But the relative error in the backscattering region III (error/mass
loss in region III) is by a factor of 10 larger. Those errors are the
errors made by approximating null infinity by grid which ends at a
finite physical distance. There is almost no difference between
$m_{\rm Haw, fin}$ and $m_{\rm Bondi, fin}$.
\section{Summary}
In the work reported in this paper a mathematical formalism, designed
to investigate the structure of a spacetime in the large by
conformal techniques, has been introduced into numerics. This paper
demonstrates that global  properties of a spacetime are numerically 
calculable and must not necessarily be estimated from a numerical
simulation by heuristic arguments. 
\\
The set of equations can be solved either by a spacelike or a 
characteristic initial value problem. The form of the
equations differs from Einstein's equations. The
technical tricks developed for the numerical 
solution of Einstein's equations cannot be used unchanged. The accuracy
problem as described in section 
\ref{AccuracyProblem} shows that 
there is still the potential for further progress. Recently, too late
to be included in this paper, a new way of discretizing was found,
which propagates the regularity condition at the center without
switching schemes near the center, as done here. The accuracy has
improved by a factor of $10$ for models near the 
critical parameter.  
\\
In this formalism important properties of spacetimes are
automatically fulfilled, the ADM mass is always conserved, the
energy radiated away equals the change in the Bondi mass, and there is no
spurious ingoing wave caused by an unphysical reflection at the outer
boundary. Furthermore the 
location of event horizons is straightforward and the calculation
yields ``Penrose'' diagrams allowing to read off the
causal structure of singularities.
\\
The obvious advantages might not be worth the effort for the
investigation of 
astrophysical questions where most of the uncertainty originates in
an approximate equation of state. 
But if one is interested in the use of 
numerical  relativity as a substitute for
the lacking experimental approach to the open problems of
mathematical relativity, the conformal techniques provide a very
promising approach. In this 
circumstance it should be mentioned that H.~Friedrich has developed
the conformal techniques further \cite{FrXXee}. There the rescaling
factor $\Omega{}$ is no longer a variable of the system fixed in a
complicated way by the gauge source function $R(t,r)$. The gauge
freedom in the rescaling is fixed directly by specifying
$\Omega(t,r)$, and the problem with spacelike infinity $i^0$ may be
solved \cite{FrXXPC}. 
\\[2cm]
It is a pleasure for me to thank the relativity group in Garching for
the cooperation during my Ph.~D.\ thesis, J. Winicour for his advice
on my first steps in numerical relativity during his stay in Garching,
and W. Kley for helpful comments on the manuscript. 
\begin{appendix}
\section{Notation}
\label{Konventionen}
The signature of the Lorentzian metric $g_{ab}$ is $(-,+,+,+)$. 
\\
Whenever possible I use abstract indices as described
in~\cite[chapter~2]{PeR84SA}. Small Latin letters denote abstract
indices, underlined  
small Latin letters are frame indices. For the components of a tensor with
respect to coordinates small Greek letters are used. The frame
$\left(\frac{\partial}{\partial x^\mu}\right)^a$ is constructed from
the coordinates $x^\mu $, $e_{\f{i}}{}^a$ denotes an arbitrary frame.
In this notation $v_a$ is a covector, $v_{\f{i}}$ a scalar, namely 
$v_a\, e_{\f{i}}{}^a$. 
\\
$v(f)$ is defined to be the action of the vector $v^a$ on the function
$f$, i.e.\ for every covariant derivative $\nabla_a$: 
$t(f)=t^a\,\nabla_a f$. 
\\
The transformation between abstract, coordinate, and frame indices is
done by contracting with $e_{\f{i}}{}^a$ and $e_{\f{i}}{}^\mu $. All
indices may be raised and lowered with the metric $g_{AB}$ and the
inverse $g^{AB}$. $g^{AC}\,g_{CB} = \delta^A{}_B$, $A$ and $B$ are
arbitrary indices, e.g.\  $e_{\f{i}a}=g_{ab}\,e_{\f{i}}{}^b$ and
$e^{\f{i}}{}_a=g^{\f{i}\f{j}}\,e_{\f{i}a}$.  
\\
For a frame $e_{\f{i}}{}^a$ and a covariant derivative 
$\nabla_a$ the Ricci rotation coefficients
are defined as 
\begin{equation*}
  \gamma^a{}_{\f{i}\f{j}} := e_{\f{i}}{}^b \nabla_b e_{\f{j}}{}^a.
\end{equation*}
From this definition follows
\begin{equation*}
  e_{\f{i}}{}^a \, e^{\f{j}}{}_b \, (\nabla_a t^b) = 
  e_{\f{i}}(t^{\f{j}}) + \gamma^{\f{j}}{}_{\f{i}\f{k}} \, t^{\f{k}}.
\end{equation*}
With respect to a coordinate frame $e_{\mu}{}^a \equiv
\left(\frac{\partial}{\partial x^\mu}\right)^a$  the components 
$\gamma^\lambda{}_{\mu\nu}$ are the Christoffel symbols
$\Gamma^\lambda{}_{\mu\nu}$.  
\\
The torsion $T^a{}_{bc}$ is defined by
\begin{equation*}
  \nabla_a\nabla_b f - \nabla_b\nabla_a f = - T^c{}_{ab} \, \nabla_c f,
\end{equation*}
the Riemann tensor $R_{abc}{}^d$ by 
\begin{equation*}
  \nabla_a\nabla_b\omega_c - \nabla_b\nabla_a\omega_c =
  R_{abc}{}^d \, \omega_d  - T^{d}{}_{ab} \, \nabla_d \omega_c.
\end{equation*}
Contraction gives the Ricci tensor,
\begin{equation*}
  R_{ab} = R_{acb}{}^c,
\end{equation*}
and the Ricci scalar
\begin{equation*}
  R = R_{ab}\, g^{ab}.
\end{equation*}
The Einstein tensor is given by
\begin{equation*}
  G_{ab} = R_{ab} - \frac{1}{2} \, R \, g_{ab}.
\end{equation*}
The speed of light $c$ is set to $1$. The gravitational constant
$\kappa $ in $G_{ab}=\kappa\, T_{ab}$ has been set to $1$ or $0$ in
the calculations, corresponding to the full nonlinear theory or a
scalar field on a flat background respectively.
\section{The system of equations in double null coordinates}
\label{AppsphSys}
\subsection{The time evolution equations}
From the system (\ref{quaSys}) the following set of equations can be
derived, where the abbreviations 
\begin{eqnarray}
  && \gamP = \gamma^{\f{0}}{}_{\f{0}\f{1}} +
       \gamma^{\f{0}}{}_{\f{1}\f{1}}                \nonumber \\
  && \gamii = \gamma^{\f{0}}{}_{\f{2}\f{2}} \nonumber \\
  &&  \gamM = \gamma^{\f{0}}{}_{\f{1}\f{1}} -
       \gamma^{\f{0}}{}_{\f{0}\f{1}}                \nonumber \\
  && \hRi = ( \hR_{\f{0}\f{1}} + \hR_{\f{1}\f{1}} )/2      \nonumber \\
  && \hRii = \hR_{\f{0}\f{0}} - \hR_{\f{1}\f{1}}           \nonumber \\
  && \hRiii = ( \hR_{\f{1}\f{1}} - \hR_{\f{0}\f{1}} )/2    \nonumber \\
  && \hphi = ( \hp_{\f{0}\f{1}} + \hp_{\f{1}\f{1}} )/2    \nonumber \\  
  && \hphiii = ( \hp_{\f{1}\f{1}} - \hp_{\f{0}\f{1}} )/2  
\end{eqnarray}
and the substitutions 
\begin{displaymath}
  \phi_a{}^a \rightarrow \frac{R}{6},
\end{displaymath}
\begin{displaymath}
  \omega \rightarrow \frac{1}{4} \, \Omega \, \RII -
     \frac{1}{4} \, \Omega^3 \, ( \hT_{\f{0}\f{0}} - \hT_{\f{1}\f{1}} )
     - \gamii \, \Omo
     - \frac{\gamma}{r} \, \Omi,      
\end{displaymath}
and
\begin{equation}
\label{hphi2}
  \phi2 = \hp_{\f{0}\f{0}} - \hp_{\f{1}\f{1}} 
        \rightarrow - 2 \, \gamii \, \pho
          - 2 \, \frac{\gamma}{r} \, \phI
          - \frac{1}{2} \, \phi_a{}^a,
\end{equation}
which follow from spherical symmetry, have been used.
\begin{eqnarray}
\label{sphSys}
\lefteqn{
    \left( \partial_{\f{0}} + \partial_{\f{1}} \right) \gamM \quad = \quad
    \mbox{} - (\gamM \, \gamP) + \dioio \, \Omega 
    + \frac{1}{2} \,  ( \frac{\RICCI}{6} - \hRii )
    }
\\ \noalign{} \nonumber \\
    \lefteqn{ 
      \left( \partial_{\f{0}} + \partial_{\f{1}} \right)
      \hRiii \quad = \quad 
      \Big( \dioio \, \Omo + \Omega \,  m  +  S_{\hRiii} \Big) / \NENNER 
      }
    \nonumber \\ & & \quad
     \mbox{} + 
      \big( - \gamP \,  ( \frac{1}{2} \,  \hRii + 2 \, \hRiii ) 
        + \gamii \,  ( \hRi  - \hRiii )
        + \frac{1}{8} \DiR - \frac{1}{24} \, \DoR
       \big)
\\ 
\noalign{with}
\lefteqn{ \quad
  S_{\hRiii} =
  \left\{  \begin{array}{cl}
    \left. \begin{array}{l}
      \gamma \,  
      \Big( \, \NENNER \,  ( \hRi + \hRiii 
                  - \frac{1}{2} \, \hRii ) / r  \\ 
      \qquad \mbox{} - \frac{\kappa}{2} \, \Omega \,  (\phI/r) \,  
            ( \Omo\, \phi + \pho\, \Omega ) \Big)
    \end{array} \right\} & \qquad \mbox{for $r\ne 0$} \\ \\
    \mbox{} - \frac{\kappa}{2} \, \gamma \,  \Omega \,    
            ( \Omo\, \phi + \pho\, \Omega ) \, \partial_r \phI 
            & \qquad \mbox{for $r=0$}
    \end{array} \right.
}
\\ \noalign{} \nonumber \\
  \lefteqn{ 
   \left( \partial_{\f{0}} + \partial_{\f{1}} \right) \hphiii \quad = \quad
      \gamii \,  ( \hphi - \hphiii )    
      - 2 \, \gamP  \, \hphiii 
      }    
    \nonumber \\ & & \quad
     \mbox{} + \frac{\pho}{2} \,  
        \left( - ( \hRi - \hRiii )- \dioio \, \Omega  
          + \frac{1}{2} \,  \hRii - \frac{1}{8} \, \RICCI + 2 \,
          \gamii \, \gamP \right) 
    \nonumber \\ & & \quad      
      \mbox{} 
      + \frac{\phI}{2} \,  ( \hRi + \hRiii + \frac{3}{8}\, \RICCI ) 
      + \phi \,  \left( \frac{1}{16} \, \DiR + \frac{1}{24} \, \gamP \,
                                  \RICCI - \frac{1}{48} \, \DoR \right)
      + S_{\hphiii}
\\
\lefteqn{ \quad
      S_{\hphiii} = 
  \left\{  \begin{array}{cl}
    \left. \begin{array}{l}
        \gamma \, 
          \Big(  \left(  (\phI / r) \,  \gamma 
                                  + \gamii \,  \pho 
                                  + \frac{1}{24} \, \phi \,  \RICCI
                                  + \hphi + \hphiii \right) / r  \\
          \qquad \mbox{} + \gamP \,  (\phI /r ) 
          \Big)
        \end{array} \right\} & \qquad \mbox{for $r\ne 0$} \\ \\
        \gamma \, \gamP \,  \partial_r \phI  & \qquad \mbox{for $r=0$}
      \end{array} \right.
}
\nonumber \\ \noalign{} \nonumber \\
\lefteqn{
    \left( \partial_{\f{0}} - \partial_{\f{1}} \right) \gamP \quad = \quad
     \mbox{} - (\gamM\, \gamP) + \dioio \, \Omega 
     + \frac{1}{2} \,  ( \frac{\RICCI}{6} - \hRii )
}
\\ \noalign{} \nonumber \\
   \lefteqn{ 
    \left( \partial_{\f{0}} - \partial_{\f{1}} \right) \hRi \quad = \quad
    \Big( \dioio \, \Omo + \Omega \,  m  + S_{\hRi} \Big)  /  \NENNER 
    }
    \nonumber \\ & & \quad
    \mbox{} +  
      \big( - \gamM \,  ( \frac{1}{2} \,  \hRii + 2 \, \hRi ) 
        - \gamii \,  ( \hRi - \hRiii ) 
        - \frac{1}{8} \, \DiR - \frac{1}{24} \, \DoR
       \big) 
\\
\lefteqn{ \quad
      S_{\hRi} =
   \left\{  \begin{array}{cl}
    \left. \begin{array}{l}
        \gamma \,  
        \Big( \mbox{} - \NENNER \,  ( \hRi + \hRiii - \frac{1}{2} \,  
\hRii ) / r \\
         \qquad \mbox{} - \frac{\kappa}{2} \,  \Omega \,  ( \phI / r ) \,  
               ( \Omo\, \phi + \pho\, \Omega ) 
         \Big)
    \end{array} \right\} & \qquad \mbox{for $r\ne 0$} \\ \\
    \mbox{} - \frac{\kappa}{2} \, \gamma \,  \Omega \,    
            ( \Omo\, \phi + \pho\, \Omega ) \, \partial_r \phI 
            & \qquad \mbox{for $r=0$}
\end{array} \right.
}
\nonumber \\ \noalign{} \nonumber \\
  \lefteqn{ 
    \left( \partial_{\f{0}} - \partial_{\f{1}} \right) \hphi \quad = \quad
      \mbox{} - \gamii \,  ( \hphi - \hphiii )
      - 2 \,  \hphi \, \gamM
      }
    \nonumber \\ & & \quad
     \mbox{} + \frac{\pho}{2} \,  
        \left( \hRi - \hRiii - \dioio \, \Omega  
          + \frac{1}{2} \,  \hRii - \frac{1}{8}
          \, \RICCI + 2 \, \gamM\, \gamii  \right) 
    \nonumber \\ & & \quad      
      \mbox{} 
      + \frac{\phI}{2} \,  
        \left( - ( \hRi + \hRiii ) - \frac{3}{8} \, \RICCI \right)
      + \phi \,  
        \left( - \frac{1}{16} \, \DiR  
          + \frac{1}{24} \gamM \, \RICCI - \frac{1}{48} \, \DoR \right)
      + S_{\hphi}
\\
\lefteqn{ \quad
      S_{\hphi} =
  \left\{  \begin{array}{cl}
    \left. \begin{array}{l}
        \gamma \, 
        \Big( - \left(  (\phI / r ) \,  \gamma
                        + \gamii \,  \pho 
                        + \frac{1}{24} \, \phi \,  \RICCI
                        + \hphi + \hphiii \right) / r  \\
         \qquad   + (\phI / r) \,  \gamM                   
         \Big)
        \end{array} \right\} & \qquad \mbox{for $r\ne 0$} \\ \\
        \gamma \, \gamP \,  \partial_r \phI  & \qquad \mbox{for $r=0$}
      \end{array} \right.
}
\nonumber \\ \noalign{} \nonumber \\
\lefteqn{
    \partial_{\f{0}} \ei \quad = \quad
      \mbox{} - \frac{1}{2} \, \ei \, ( \gamM + \gamP )
}
\\ \noalign{} \nonumber \\
\lefteqn{
   \partial_{\f{0}} e = \mbox{} - e \,  \gamii
}
\\ \noalign{} \nonumber \\
\lefteqn{
    \partial_{\f{0}} \gamii \quad = \quad \mbox{}
      - \left(\gamii \right)^2 - \frac{1}{2} \,  \dioio  \,  \Omega 
      + \frac{1}{2} \,  
      \left( - ( \hRi + \hRiii ) - \frac{1}{2} \,  \hRii 
+ \frac{\RICCI}{6} \right) 
      + S_{\gamii}
}
\\   
\lefteqn{ \quad
      S_{\gamii} =
  \left\{  \begin{array}{cl}
    \begin{array}{l}
        \frac{\gamma}{2} \,  ( \gamM - \gamP) / r 
    \end{array} & \qquad \mbox{for $r\ne 0$} \\ \\
    \frac{\gamma}{2} \, \partial_r ( \gamM - \gamP ) & 
\qquad \mbox{for $r=0$}
 \end{array} \right.
}
\nonumber \\ \noalign{} \nonumber \\
\lefteqn{
    \partial_{\f{0}} \gamma \quad = \quad
    \mbox{} - \gamii\, \gamma
    + \frac{r}{2} \,  \left( (\gamM - \gamP)\, \gamii 
+ (\hRi - \hRiii) \right)
}
\\ \noalign{} \nonumber \\
    \lefteqn{ 
      \partial_{\f{0}} \hRii \quad = \quad
      \big( \mbox{} - 2 \,  ( \Omega \,  m + \dioio  \, \Omo ) 
                + S_{\hRii} \big) / \NENNER
      }
    \nonumber \\ & & \quad
    \mbox{} + \big( \gamii \,  ( - 2 \,  ( \hRi + \hRiii )  - 3 \,  \hRii )
                       - \frac{1}{6} \,  \DoR \big) 
    \hspace{5cm}
\\
\lefteqn{ \quad
      S_{\hRii} =
  \left\{  \begin{array}{cl}
    \left. \begin{array}{l}
        \gamma \, 
        \Big( \mbox{} - 2 \,  \NENNER \,  ( \hRi - \hRiii ) / r  \\
          \qquad \mbox{} + \kappa\,  \Omega \,  (\phI / r) \,  
( \Omo\, \phi + \pho\, \Omega )
         \Big)
    \end{array} \right\} & \qquad \mbox{for $r\ne 0$} \\ \\
    \left. \begin{array}{l}
        \gamma \, 
        \Big( \mbox{} - 2 \,  \NENNER \,  \partial_r ( \hRi - \hRiii ) \\
          \qquad \mbox{} + \kappa\,  \Omega \,  ( \Omo \, \phi + \pho \,
                                                              \Omega )
                                                    \, \partial_r \phI 
         \Big)    
       \end{array} \right\} & \qquad \mbox{for $r = 0$}
  \end{array} \right.
}
\nonumber \\ \noalign{} \nonumber \\
\lefteqn{
    \partial_{\f{0}} \Omega \quad = \quad \Omo
}
\\ \noalign{} \nonumber \\ 
    \lefteqn{ 
      \partial_{\f{0}} \Omo \quad = \quad      
      \gamii\, \Omo - \frac{1}{2} \, ( \gamM - \gamP ) \,  \Omi 
      }
    \nonumber \\ & & \quad
     \mbox{} + 
      \Omega \,  
        \bigg( 
          \frac{\kappa}{4} \,  \Omega^2 \, 
            \Big( \phi \,  \big( \frac{\RICCI}{8} \, \phi 
- ( \hphi + \hphiii ) \big)
              + \pho \,  ( 3 \, \gamii\, \phi + 2 \, \pho ) 
     \nonumber \\ & & \quad   
     \hphantom{\mbox{} + \Omega \,  \Big( } \quad
     \mbox{}  
     - \frac{1}{2} \,  \left( ( \hRi + \hRiii ) 
+ \frac{3}{2} \, \hRii \right) \,  \NENNER
          \Big)
         \bigg)
    \nonumber \\ & & \quad
     \mbox{}  + S_{\Omo}
\\
\lefteqn{ \quad
      S_{\Omo} =
  \left\{  \begin{array}{cl}
    \begin{array}{l}
        \gamma \,  \Big( \Omi/r + \frac{3}{4} \, \kappa\, \Omega^3 \,  
                                             \phi \,  (\phI / r) \Big)    
    \end{array} & \qquad \mbox{for $r\ne 0$} \\ \\
    \begin{array}{l}
        \gamma \,  \Big( \partial_r \Omi 
                                    + \frac{3}{4} \, \kappa\, \Omega^3 \,  
                                    \phi \,  \partial_r \phI \Big)    
    \end{array} & \qquad \mbox{for $r = 0$}
  \end{array} \right.
}
\nonumber \\ \noalign{} \nonumber \\
  \lefteqn{
    \partial_{\f{0}} \Omi \quad = \quad \mbox{}
    - \frac{1}{2}  \, \NENNER \,  ( \hRi - \hRiii ) \,  \Omega 
      }
    \nonumber \\ & & \quad   
    \mbox{}  + 
      \frac{\kappa}{2} \, \Omega^3 \,  
      \left( \pho\, \phI - \frac{1}{2} \, ( \hphi - \hphiii ) \,  
\phi \right)  
      - \frac{1}{2} \, ( \gamM - \gamP ) \,  \Omo 
\\ \noalign{} \nonumber \\
\lefteqn{
  \partial_{\f{0}} \phi \quad = \quad \pho
}
\\ \noalign{} \nonumber \\
\lefteqn{
    \partial_{\f{0}} \pho \quad = \quad 
      \mbox{} 
      - 2 \,  \gamii \,  \pho 
      - \frac{\phI}{2} \,  ( \gamM - \gamP ) 
      - \phi\, \frac{\RICCI}{8} 
      + \hphi + \hphiii  
      + S_{\pho}
}
\\ \nonumber \\
\lefteqn{ \quad
      S_{\pho} = 
  \left\{  \begin{array}{cl}
    \begin{array}{l}
      \mbox{} - 2 \,  \gamma \,  (\phI / r)
    \end{array} & \qquad \mbox{for $r\ne 0$} \\ \\
    \begin{array}{l}
      \mbox{} - 2 \,  \gamma \,  \partial_r \phI
    \end{array} & \qquad \mbox{for $r = 0$}
  \end{array} \right.
}
\\ \noalign{} \nonumber \\
\lefteqn{
    \partial_{\f{0}} \phI \quad = \quad
    \hphi - \hphiii 
      - \frac{\pho}{2} \, ( \gamM - \gamP )
}
\nonumber \\ \noalign{} \nonumber \\
\lefteqn{
    \partial_{\f{0}} \dioio \quad = \quad
    \mbox{} - 3 \,  \gamii \, \dioio  + 
    \left( m + \frac{\kappa}{4} \, \phi^2 \,  \Omo \,  \Omega \, \dioio
      + S_{\dioio}  \right) / \NENNER
}
\\ 
\lefteqn{ \quad
     S_{\dioio} =
  \left\{  \begin{array}{cl}
    \begin{array}{l}
        \mbox{} - \gamma \, \frac{\kappa}{2} \, (\phI / r) \, 
        ( \Omo \,  \phi + \Omega \,  \pho )
    \end{array} & \qquad \mbox{for $r\ne 0$} \\ \\
    \begin{array}{l}        
      \mbox{} - \gamma \, \frac{\kappa}{2} \,
       ( \Omo \,  \phi + \Omega \,  \pho ) \, \partial_r \phI
    \end{array} & \qquad \mbox{for $r = 0$}
  \end{array} \right.
} \nonumber
\end{eqnarray}
where $m$ stands for
\begin{eqnarray*}
\lefteqn{m \quad = \quad \frac{\kappa}{2} \,  
      ( \phi \,  \Omi + \phI \,  \Omega ) \,  ( \hphi - \hphiii ) }
    \nonumber \\ & & \quad \mbox{}
      - \frac{\kappa}{2} \,  ( \phi \,  \Omo + \pho \,  \Omega ) \,  
( \hphi + \hphiii )  
      + \frac{\kappa}{4} \,  \phI \, 
        \left( 4 \,  \Omo \,  \phI
          - \phi \,  \Omega \,  ( \hRi - \hRiii ) 
          \right) 
    \nonumber \\ & & \quad \mbox{}
      + \kappa\,  \pho \,  
        \Bigg( \frac{1}{4} \, \phi \,  
          \bigg( - 2 \,  \gamii\, \Omo
            + \Omega \,  
              \left( \dioio \, \Omega + (\hRi + \hRiii - \frac{1}{2} \hRii) 
                - \frac{1}{12} \, \RICCI
               \right) \bigg)  
    \nonumber \\ & & \quad \hphantom{\bigg(
             \frac{1}{4} \, \phi \, } \quad \mbox{}
          - \frac{1}{2} \,  \gamii \,  \Omega\, \pho
          - \Omi\, \phI          
         \Bigg) 
    \nonumber \\ & & \quad \mbox{}
      + \frac{\kappa}{4} \,  \phi^2 \, 
        \bigg( - \Omi \,  ( \hRi - \hRiii )
          + \Omo \,  \left(  ( \hRi + \hRiii - \frac{1}{2} \,  \hRii ) 
                                        - \frac{1}{12} \, \RICCI \right)
         \bigg)
\end{eqnarray*}
\subsection{The Constraints}
From (\ref{quaSys}) the following set of constraint equations can be
derived:
\begin{eqnarray}
\lefteqn{
    \partial_{\f{1}} \ei \quad = \quad 
      \mbox{} - \frac{1}{2} \, ( \gamM - \gamP )
      }
\\ \nonumber \\
\lefteqn{
    \partial_{\f{1}} e \quad = \quad 
    e \, \frac{\ei + \gamma}{r} \quad \mbox{for $r \ne 0$, $0$ otherwise}
    }
\\ \nonumber \\
\lefteqn{
    \partial_{\f{1}} \gamii \quad = \quad 
    \frac{1}{2} \, \left( \hRiii - \hRi \right)
    + \frac{\gamma}{r} \, \left( \gamii - \frac{1}{2} \, ( \gamP +
    \gamM ) \right)
}
\nonumber \\ & & \quad
\mbox{for $r \ne 0$, $0$ otherwise}
\\ \nonumber \\
\lefteqn{
    \partial_{\f{1}} \gamma \quad = \quad 
    \gamma \, \frac{\ei + \gamma}{r}
    }
    \nonumber \\ & & \quad
    \mbox{} + \frac{r}{2} \,     
    \left( \frac{\RICCI}{6} + \hRi + \frac{\hRii}{2} 
                  + \hRiii - \dioio \, \Omega
                  - ( \gamP + \gamM ) \, \gamii 
                \right)
\nonumber \\ & & \quad
\mbox{for $r \ne 0$, $0$ otherwise}
\\ \nonumber \\
\label{ND1hR2}
\lefteqn{
    \partial_{\f{1}} \hRii \quad = \quad 
    2 \, \gamii \, \left( \hRiii - \hRi \right) - \frac{1}{6} \, \DiR
    }
    \nonumber \\ & & \quad
    \mbox{} + \left( - 2 \, ( \Omega \, m_C + \dioio \, \Omi ) 
                                + s_{\hRii} \right)
    / \NENNER, 
\nonumber \\ 
& & \qquad s_{\hRii} =
        \gamma \, 
        \Big( - 2 \,  \NENNER \,  
                  ( \hRi + \hRiii - \frac{1}{2} \, \hRii ) /r  
\nonumber \\ & & \qquad \qquad \quad
          \qquad \mbox{} + 3 \, \kappa\,  \Omega \,  (\phI / r) \,  
                                           ( \Omi\, \phi + \phI\, \Omega )
         \Big)
\nonumber \\ & & \quad
\mbox{for $r \ne 0$, $0$ otherwise}
\\ \nonumber \\
\lefteqn{
    \partial_{\f{1}} \Omega \quad = \quad \Omi
      }
\\ \nonumber \\
\lefteqn{
    \partial_{\f{1}} \Omo \quad = \quad
      \Omi \, \frac{\gamM + \gamP}{2} - \frac{\hRi-\hRiii}{2} \, \Omega
    }
    \nonumber \\ & & \quad
    \mbox{} + \frac{\kappa}{4} \, \Omega^3 \,
        \left( 2 \, \pho \, \phI + \phi \, ( \hphiii - \hphi ) + 
        \frac{\phi^2}{2} ( \hRi - \hRiii )
        \right)
\\ \nonumber \\
\label{ND1Om}
\lefteqn{
    \partial_{\f{1}} \Omi \quad = \quad
      \left( \frac{ \gamP + \gamM }{2} - \gamii \right) \, \Omo 
      - \frac{\gamma}{r} \, \Omi 
      - \frac{\Omega}{2} \left( \hRi - \frac{\hRii}{2} + \hRiii \right)
    }
    \nonumber \\ & & \quad
    \mbox{} + \frac{\kappa}{4} \, \Omega^3 \, 
       \left( 2 \, \phI^2 + \phi \, \left( \hphiii - \hphi 
                      - \gamii \, \pho - \frac{\gamma}{r} \, \phI 
                      + \frac{\phi}{2} \, ( \hRi - \frac{\hRii}{2} +
                      \hRiii - \frac{\RICCI}{12} ) \right)
       \right)
\nonumber \\ & & \quad
\mbox{for $r \ne 0$}
\\ \nonumber \\
%
\lefteqn{
    \partial_{\f{1}} \phi \quad = \quad \phI
     }
\\ \nonumber \\
\lefteqn{
    \partial_{\f{1}} \pho \quad = \quad 
        \hphi - \hphiii + \frac{\gamM  + \gamP}{2} \, \phI
     }
\\ \nonumber \\
\lefteqn{
    \partial_{\f{1}} \phI \quad = \quad 
      \hphi + \hphiii + \frac{\gamM  + \gamP}{2} \, \pho
      + \frac{\RICCI}{24} \, \phi
      }
\\ \nonumber \\
\label{ND1d}
\lefteqn{
    \partial_{\f{1}} \dioio \quad =  \quad
    \left( m_C + \frac{\kappa}{4} \, \phi^2 \,  \Omi \,  \Omega \, \dioio
      + s_{\dioio}  \right) / \NENNER
    }
\\ 
& & \qquad s_{\dioio} =
        \gamma \, 
        \Bigg( 3 \, \gamma \, 
        \Big( \NENNER \, \dioio/r 
\nonumber \\ & & \qquad \qquad \quad
          \qquad \mbox{} - \frac{\kappa}{2} \, \phI/r \, \left( \Omega \,
                    \phI + \Omi \, \phi \right) \Big)
         \Bigg)
\nonumber \\ & & \quad
\mbox{for $r \ne 0$}
\\ \nonumber \\
\noalign{where $m_C$ stands for}
\lefteqn{m_C \quad = \quad \frac{\kappa}{2} \, 
      ( \phi \, \Omi + \phI \, \Omega ) \, ( \hphi + \hphiii ) 
    - \frac{\kappa}{2} \,  ( \phi \, \Omo + \pho \, \Omega ) \, 
       ( \hphi - \hphiii ) }
    \nonumber \\ & & \quad \mbox{}
    + \frac{\kappa}{4} \, \pho \, 
        \left( 4 \, \Omo \, \phI - 4 \, \Omi \, \pho 
                    - 6 \, \gamii \, \Omi \, \phi
          + \phi \,  \Omega \,  ( \hRi - \hRiii ) 
          \right) 
    \nonumber \\ & & \quad \mbox{}
      + \kappa\,  \phI \,  
        \Bigg(  \frac{1}{4} \, \phi \, \Omega \,
          \left( \dioio \, \Omega - (\hRi + \hRiii - \frac{3}{2} \hRii) 
                - \frac{1}{4} \, \RICCI
               \right)  - 6 \, \gamii \, \Omega \, \pho
    \nonumber \\ & & \quad \mbox{}
      + \frac{\kappa}{4} \phi^2 \,  
        \bigg( \Omo \, ( \hRi - \hRiii )  
        - \Omi \, \left( \hRi + \hRiii + \frac{3}{2} \hRii) 
                + \frac{1}{4} \, \RICCI \right)  \bigg).
    \nonumber \\ & & \quad \mbox{}
\end{eqnarray}
Equation (\ref{SpGl}), which must be fulfilled at one point, becomes
\begin{eqnarray}
\label{SingGlSph}
&&
  R\,{{\Omega}^2} + 6\,\hRii\,{{\Omega}^2} - 
  24\,\gamii\,\Omega\,\Omo
  + 12\,{\Omo^2} - 
  24\,\frac{\gamma}{r}\,\Omega\,\Omi
  - 12\,{{\Omi}^2} - 
  {{\frac{\kappa}{4}\,R\,{{\Omega}^4}\,{{\phi}^2}}} 
  \nonumber \\ &&
  - {{\frac{3}{2}\,\kappa\,\hRii\,{{\Omega}^4}\,{{\phi}^2}}} - 
  6\,\kappa\,\gamii\,{{\Omega}^4}\,\phi\,\pho - 
  3\,\kappa\,{{\Omega}^4}\,{{\pho}^2} - 
  6\,\kappa\,\frac{\gamma}{r}\,{{\Omega}^4}\,\phi\,\phI
  \nonumber \\ &&
  + 3\,\kappa\,{{\Omega}^4}\,{{\phI}^2}
  \qquad = \qquad 0.
\end{eqnarray}
Because of spherical symmetry there is also the identity 
\begin{equation}
\label{dSubst}
  \Omega \, \dioio = 
  \frac{e^2 - \gamma^2 }{r^2} + \left(\gamii \right)^2 
   - \frac{1}{2} \, \hRii - \frac{\RICCI}{12}.
\end{equation}
\subsection{The regularity conditions}
\label{regularityConditions} 
At the center the following must hold for a regular spacetime:
\begin{mathletters}
\begin{eqnarray}
 e & = & - \gamma = e_{\f{1}}{}^{\f{1}} \\
 \gamM & = & \gamP \\
 \gamii & = & \frac{1}{2} \, \left(\gamM + \gamP \right) \\
 \hRi & = & \hRiii \\
 \hRii & = & 2\,\left( \hRi+\hRiii \right) \\
 \dioio & = & 0 \\
 \Omi & = & 0 \\
 \phI& = & 0 \\
 \hphi & = & \hphiii.
\end{eqnarray}
On \Scri{} 
\begin{eqnarray}
  \label{NullScri}
  \Omo -\Omi & = & 0 \\
  \label{smoothScri}
  \hRi + \hRiii & = & \frac{1}{\Omi} \, \Bigg( 
           - \partial_{\f{1}}\partial_{\f{1}}\Omi
           - \frac{\gamma}{r} \, \partial_{\f{1}}\Omi
           + \frac{\gamM + \gamP}{2} \, \left(  \frac{\gamM + \gamP}{2} 
                                                    - \gamii \right)
\nonumber \\
    & & \quad \qquad
   \mbox{} + \frac{\Omo}{2} \, ( \hRi - \hRiii )  - \Omi \, \frac{R}{12}
    + \Omi \,  \left( \frac{\gamM + \gamP}{2} \, \gamii 
                                - (\frac{\gamma}{r})^2 \right)
\nonumber \\
     & & \quad \qquad
     \mbox{} + \Omo \, \left( \partial_{\f{1}} \, \frac{\gamM + \gamP}{2} 
                                + \left(  \frac{\gamM + \gamP}{2}  
                                                - \gamii \, \right) \,
                                                \frac{\gamma}{r}
                                              \right)
     \Bigg).
\end{eqnarray}
\end{mathletters}
Equation (\ref{NullScri}) also guarantees that equation
(\ref{SingGlSph}) is fulfilled on at least one point in $M$, namely
\Scri{}.  
\end{appendix}

\end{document}